\documentclass[amsmath,amssymb,aps,floats,amsfonts,notitlepage,superscriptaddress,nofootinbib,prd,a4paper,twocolumn,longbibliography]{revtex4-1}

\usepackage[utf8]{inputenc}
\usepackage[]{graphicx}
\usepackage{hyperref}
\usepackage{url}
\usepackage{color}
\usepackage[usenames,dvipsnames,svgnames,table]{xcolor}
\usepackage{multirow}
\usepackage{ulem}
\usepackage[scr=boondoxo,scrscaled=1.05]{mathalfa}%script fonts

\font\ec=ecrm0800 at 11pt
\def\th{\hbox{\ec\char'336}}
\def\edth{\hbox{\ec\char'360}}

\newcommand{\emm}{{\mathscr m}}

\newcommand{\msout}[1]{}

\newcommand{\mb}{{\bar{m}}}

\begin{document}

\title{Separable electromagnetic perturbations of rotating black holes}

\author{Barry Wardell}
\affiliation{
School of Mathematics and Statistics,
University College Dublin,
Belfield,
Dublin 4,
D04 V1W8,
Ireland
}

\author{Chris Kavanagh}
\affiliation{
Max Planck Institute for Gravitational Physics (Albert Einstein Institute),
Am M\"uhlenberg 1,
Potsdam 14476,
Germany}

\begin{abstract}
We identify a set of Hertz potentials for solutions to the vector wave equation
on black hole spacetimes. The Hertz potentials yield Lorenz gauge electromagnetic vector
potentials that represent physical solutions to the Maxwell equations, satisfy the Teukolsky
equation, and are related to the Maxwell scalars by straightforward and separable inversion relations.
Our construction, based on the GHP formalism, avoids the need for a mode ansatz and
leads to potentials that represent both static and non-static solutions. 
As an explicit example, we specialise the procedure to mode-decomposed
perturbations of Kerr spacetime and in the process make connections with previous results.
\end{abstract}

\maketitle

\section{Introduction}
\label{sec:introduction}

In a seminal work, Teukolsky \cite{Teukolsky:1972my} showed that the equations
governing perturbations of rotating black holes can be recast into a form where
they are given by decoupled equations. These equations further had the
remarkable property of being separable, reducing the problem to the solution of
a set of uncoupled ordinary differential equations. In the electromagnetic case,
Teukolsky's results yield solutions for the spin-weight $\pm1$ components of the
Faraday tensor, but do not give a method for obtaining a corresponding vector
potential. Subsequent results (and their corresponding equivalents for
gravitational perturbations) 
\cite{Chrzanowski:1975wv,Kegeles:1979an,Wald:1978vm,Whiting:2005hr,Pound:2013faa,Stewart:1978tm}
derived a method for reconstructing a vector potential from a Hertz potential, which in turn can be
obtained from the spin-weight $\pm1$ components of the Faraday tensor.
These were initially restricted to the radiation gauge, but have recently been
extended to the Lorenz gauge case
\cite{Sago:2002fe,Lunin:2017drx,Frolov:2018pys,Krtous:2018bvk,Frolov:2018ezx,Dolan:2018dqv,Dolan:2019hcw,Houri:2019lnu}.

In this work, we reformulate the Lorenz gauge Hertz potential of Dolan \cite{Dolan:2019hcw}
using the Geroch-Held-Penrose (GHP) formalism, which allows us to derive his results without requiring
a mode decomposition. Furthermore, our method has allowed us to identify
additional Lorenz gauge Hertz potentials which are more generally applicable.
For example, Dolan's result involved division by the frequency $\omega$,
which fails in the static $\omega=0$ case, whereas our potentials
do not have this limitation.

The layout of this paper is as follows: in Sec.~\ref{sec:EM} we review some
relevant background material, including details on the Maxwell equations in
curved spacetime, the GHP formalism, the Teukolsky equations and
Teukolsky-Starobinsky identities, and radiation gauge reconstruction of the
vector potential; in Sec.~\ref{sec:MaxwellHertz} we describe methods for
reconstructing the vector potential in Lorenz gauge; in Sec.~\ref{sec:Modes}
we give coordinate expressions for our results decomposed into spin-weighted
spheroidal harmonic modes. Finally, we provide some concluding remarks in
\ref{sec:Conclusions}.

Throughout this work we follow the conventions of Misner, Thorne and Wheeler
\cite{Misner:1974qy}: a ``mostly positive'' metric signature, $(-,+,+,+)$, is
used for the spacetime metric; the connection coefficients are defined by
$\Gamma^{\lambda}_{\mu\nu}=\frac{1}{2}g^{\lambda\sigma}(g_{\sigma\mu,\nu}
+g_{\sigma\nu,\mu}-g_{\mu\nu,\sigma})$; the Riemann tensor is
$R^{\tau}{}_{\!\lambda\mu\nu}=\Gamma^{\tau}_{\lambda\nu,\mu}
-\Gamma^{\tau}_{\lambda\mu,\nu}+\Gamma^{\tau}_{\sigma\mu}\Gamma^{\sigma}_{\lambda\nu} -\Gamma^{\tau}_{\sigma\nu}\Gamma^{\sigma}_{\lambda\mu}$, the Ricci
tensor and scalar are $R_{\mu\nu}=R^{\tau}{}_{\!\mu\tau\nu}$ and
$R=R_{\mu}{}^{\!\mu}$, and the Einstein equations are
$G_{\mu\nu}=R_{\mu\nu}-\frac{1}{2}g_{\mu\nu}R=8\pi
T_{\mu\nu}$. Standard geometrised units are used, with $c=G=1$.
We use Greek letters for spacetime indices, denote symmetrisation
of indices using round brackets [e.g. $T_{(\alpha \beta)} = \tfrac12 (T_{\alpha \beta}+T_{\beta
\alpha})$] and anti-symmetrisation using square brackets [e.g. $T_{[\alpha \beta]} = \tfrac12
(T_{\alpha \beta}-T_{\beta \alpha})$], and exclude indices from symmetrisation by surrounding them by
vertical bars [e.g. $T_{(\alpha | \beta | \gamma)} = \tfrac12 (T_{\alpha \beta \gamma}+T_{\gamma \beta
\alpha})$].

\section{Electromagnetic perturbations of type-D spacetimes}
\label{sec:EM}

\subsection{Geroch-Held-Penrose formalism}

In this work, we make use of the formalism of Geroch, Held and Penrose (GHP) \cite{Geroch:1973am},
which provides a compact means of working with tensor equations when a tetrad based on a pair of
null directions is available. Here we provide a concise review of the key features needed for this work, see Refs.~\cite{Price:Thesis,Aksteiner:2014zyp,Penrose:1987uia} for detailed reviews, and Ref.~\cite{Wardell:Pound} for a review using conventions and notation consistent with that used here.

The GHP formalism prioritises the concepts of spin- and boost-weights; within the
GHP formalism, everything has a well-defined type $\{p,q\}$, which is related to its spin-weight
$s=(p-q)/2$ and its boost-weight $b=(p+q)/2$. Only objects of the same type can be added together,
providing a useful consistency check on any equations. Multiplication of two quantities yields
a resulting object with type given by the sum of the types of its constituents.

The formalism relies on the introduction of a null tetrad $(l,n,m,\mb)$ with normalisation
\begin{equation}
  l^\alpha n_\alpha = -1, \quad m^\alpha \mb_\alpha = 1,
\end{equation}
and with all other inner products vanishing. In terms of the tetrad vectors, the metric may be written as
\begin{equation}
  g_{\alpha\beta} = -2 l_{(\alpha} n_{\beta)} + 2 m_{(\alpha} \mb_{\beta)}.
\end{equation}
There are three discrete transformations that reflect the inherent symmetry in the GHP formalism,
corresponding to simultaneous interchange of the tetrad vectors:
\begin{enumerate}
    \item $'$: $l^\alpha \leftrightarrow n^\alpha$ and $m^\alpha \leftrightarrow \bar{m}^\alpha$, $\{p,q\} \rightarrow \{-p, -q\}$;
    \item $\bar{\phantom{m}}$: $m^\alpha \leftrightarrow \bar{m}^\alpha$, $\{p,q\} \rightarrow \{q, p\}$;
    \item $\ast$: $l^\alpha \rightarrow m^\alpha$, $n^\alpha \rightarrow -\bar{m}^\alpha$, $m^\alpha \rightarrow -l^\alpha$, $\mb^\alpha \rightarrow \bar{n}^\alpha$.
\end{enumerate}
The 8 spin coefficients of well defined GHP type are defined as the directional derivatives of the tetrad vectors,
\begin{alignat}{2}
  \kappa = - l^\mu m^\nu \nabla_\mu l_\nu, & \quad \sigma = - m^\mu m^\nu \nabla_\mu l_\nu, \nonumber \\
  \rho = -\mb^\mu m^\nu \nabla_\mu l_\nu, & \quad \tau = - n^\mu m^\nu \nabla_\mu l_\nu,
\end{alignat}
along with their primed variants, $\kappa'$, $\sigma'$, $\rho'$ and $\tau'$. These have GHP type given by
\begin{alignat}{6}
  \kappa : \{3,1\}, & \quad \sigma : \{3,-1\}, & \quad \rho : \{1,1\},\quad \tau : \{1,-1\}.
\end{alignat}
The remaining spin coefficients
are used to define the GHP derivative operators (that depend on the GHP type of the object on which they are acting),
\begin{alignat}{3}
\th   &\equiv (l^\alpha \nabla_\alpha - p \epsilon - q \bar{\epsilon}), & \quad
\th'  &\equiv (n^\alpha \nabla_\alpha + p \epsilon' + q \bar{\epsilon}'),\nonumber \\
\edth  &\equiv (m^\alpha \nabla_\alpha - p \beta + q \bar{\beta}'),& \quad
\edth' &\equiv (\bar{m}^\alpha \nabla_\alpha + p \beta' - q\bar{\beta}),
\end{alignat}
where the spin coefficients are given by
\begin{subequations}
\begin{align}
  \beta &= \frac{1}{2} (m^\mu \mb^\nu \nabla_\mu m_\nu-m^\mu n^\nu \nabla_\mu l_\nu), \\
  \epsilon &= \frac{1}{2} (l^\mu \mb^\nu \nabla_\mu m_\nu-l^\mu n^\nu \nabla_\mu l_\nu),
\end{align}
\end{subequations}
along with their primed variants, $\beta'$ and $\epsilon'$. The action of a GHP derivative causes the type to change by an amount $\{p,q\}\to\{p+r,q+s\}$ where $\{r,s\}$ for each of the operators is given by
\begin{alignat}{4}
  \th : \{1,1\}, & \,\,\, \th' : \{-1,-1\}, & \,\,\, \edth : \{1,-1\},\,\,\, \edth' : \{-1,1\}.
\end{alignat}
The adjoints of the GHP operators are given by
\begin{alignat}{3}
\th^\dag   &\equiv -(\th- \rho - \bar{\rho}), & \quad
\th'^\dag  &\equiv -(\th'- \rho' - \bar{\rho}'), \nonumber \\
\edth^\dag  &\equiv -(\edth- \tau -\bar{\tau}'),& \quad
\edth'^\dag &\equiv -(\edth'- \tau' - \bar{\tau}),
\end{alignat}
and may also be written concisely as
\begin{equation}
\mathcal{D}^\dag = - (\psi_2 \bar{\psi}_2)^{1/3} \mathcal{D} (\psi_2 \bar{\psi}_2)^{-1/3}, \,\, \mathcal{D}\in\{\th, \th', \edth, \edth'\}.
\end{equation}
The Weyl scalars are defined to be the tetrad components of the Weyl tensor,
\begin{gather}
  \psi_0 = C_{lmlm},\quad
  \psi_1 = C_{lnlm}, \quad
  \psi_2 = C_{lm\mb n}, \nonumber \\
  \psi_3 = C_{ln\mb n},\quad
  \psi_4 = C_{n\mb n\mb}.
\end{gather}
These have types inherited from the tetrad vectors that appear in their definition,
\begin{gather}
  \psi_0 : \{4,0\}, \quad \psi_1 : \{2,0\}, \quad \psi_2 : \{0,0\},\nonumber \\
    \psi_3 : \{-2,0\}, \quad \psi_4 : \{-4,0\}.
\end{gather}

Many of the results that follow will be specialised to type-D spacetimes with $l^\mu$ and $n^\mu$ aligned to
the two principal null directions, in which case the Goldberg-Sachs theorem implies that 4 of the of the spin coefficients vanish,
\begin{equation}
  \kappa = \kappa' = \sigma = \sigma' = 0,
\end{equation}
and also that most of the Weyl scalars vanish
\begin{equation}
  \psi_0 = \psi_1 = \psi_3 = \psi_4 = 0.
\end{equation}

The GHP equations give relations between the Weyl scalars and the directional derivatives of the spin coefficients.
Similarly the commutator of any pair of directional derivatives can be written in terms of a linear combination of
spin coefficients multiplying single directional derivatives. Specialising to type-D spacetimes, the GHP equations
are
\begin{alignat}{4}
  \th \rho &= \rho^2, & \quad \th \tau &= \rho(\tau-\bar{\tau}'), \nonumber \\
  \edth \rho &= \tau(\rho-\bar{\rho}), & \quad \edth \tau &= \tau^2, \nonumber \\
  \th' \rho &= \rho\bar{\rho}'  - \tau \bar{\tau} &- \psi_2 &+ \edth' \tau,
\end{alignat}
the Bianchi identities are
\begin{alignat}{4}
  \th \psi_2 &= 3 \rho \psi_2, &\quad  \edth \psi_2 &= 3 \tau \psi_2,
\end{alignat}
and the commutators of the GHP operators are
\begin{subequations}
\begin{align}
  [\th, \th'] &= (\bar{\tau} - \tau')\edth + (\tau - \bar{\tau}')\edth' \nonumber \\
  & \qquad - p (\psi_2 - \tau \tau') - q (\bar{\psi}_2 - \bar{\tau}\bar{\tau}'), \\
  [\th, \edth] &= \bar{\rho}\edth - \bar{\tau}'\th + q \bar{\rho}\bar{\tau}', \\
  [\edth, \edth'] &= (\bar{\rho}' - \rho')\th + (\rho-\bar{\rho})\th'\nonumber\\
  & \qquad + p (\psi_2 + \rho \rho') - q (\bar{\rho}\bar{\rho}' + \bar{\psi}_2),
\end{align}
\end{subequations}
along with the conjugate, prime, and prime conjugate of these. 

If we further restrict to spacetimes that admit a Killing tensor, $K_{\alpha \beta} = K_{(\alpha\beta)}$, that satisfies
$\nabla_{(\alpha} K_{\beta\gamma)} = 0$,
the associated symmetries lead to additional identities relating the spin coefficients,
\begin{equation}
  \frac{\rho}{\bar{\rho}} = \frac{\rho'}{\bar{\rho}'} =  -\frac{\tau}{\bar{\tau}'}=  -\frac{\tau'}{\bar{\tau}} = \frac{\bar{C}^{1/3}}{C^{1/3}}\frac{\psi_2^{1/3}}{\bar{\psi}_2^{1/3}} = \frac{\bar{\zeta}}{\zeta},
\end{equation}
for some complex function $C$ that is annihilated by $\th$.\footnote{In the case of Kerr spacetime $C=M$ (the mass of the spacetime), so it is a real constant.}
These identities can be used along with the GHP equations to obtain a complementary set of identities,
\begin{subequations}
\begin{gather}
  \th \tau' = 2\rho \tau' = \edth' \rho, \\
  \th' \rho = \rho \rho' + \tau' (\tau - \bar{\tau}') -\frac12 \psi_2 - \frac{\bar{\zeta}}{2\zeta} \bar{\psi}_2, \\
  \edth' \tau = \tau \tau' + \rho (\rho' - \bar{\rho}') + \frac12 \psi_2 - \frac{\bar{\zeta}}{2\zeta}  \bar{\psi}_2,
\end{gather}
\end{subequations}
along with the conjugate, prime, and prime conjugate of these equations. Here, we have introduced the Killing spinor
coefficient
\begin{equation}
\zeta = -C^{1/3} \psi_2^{-1/3}, 
\end{equation}
which arises as the only non-zero spinor component of the Killing-Yano tensor. A consequence
of these additional relations is that there is an operator
\begin{equation}
  \mathcal{\pounds}_\xi = -\zeta \big( - \rho' \th + \rho \th' + \tau' \edth - \tau \edth') - \frac{p}{2} \zeta \psi_2 - \frac{q}{2} \bar{\zeta} \bar{\psi}_2,
\end{equation}
associated with the Killing vector
\begin{equation}
  \xi^\alpha = -\zeta(-\rho' l^\alpha + \rho n^\alpha + \tau' m^\alpha - \tau \mb^\alpha).
\end{equation}
There is a second operator
\begin{align}
  \mathcal{\pounds}_\eta &= -\tfrac{\zeta}{4}
    \big[(\zeta-\bar{\zeta})^2(\rho' \th - \rho \th') - (\zeta+\bar{\zeta})^2(\tau' \edth - \tau \edth') \big] \nonumber \\
      & \quad + p\, {}_\eta h_1 + q\, {}_\eta \bar{h}_1,
\end{align}
where
\begin{align}
  {}_\eta h_1 &= \tfrac18 \zeta(\zeta^2+\bar{\zeta}^2)\psi_2 - \tfrac14 \zeta\bar{\zeta}^2 \bar{\psi}_2 \nonumber \\
     & \quad + \tfrac12 \rho \rho' \zeta^2 (\bar{\zeta}-\zeta) + \tfrac12 \tau \tau' \zeta^2 (\bar{\zeta}+\zeta).
\end{align}
This is associated with the second Killing vector
\begin{equation}
  \eta^\alpha = -\tfrac{\zeta}{4}\big[(\zeta-\bar{\zeta})^2 (\rho' l^\alpha - \rho n^\alpha) -(\zeta+\bar{\zeta})^2 (\tau' m^\alpha - \tau \mb^\alpha) \big].
\end{equation}
Both $\mathcal{\pounds}_\xi$ and $\mathcal{\pounds}_\eta$ commute with all of the GHP operators and annihilate all of the spin coefficients and $\psi_2$. 
 
\subsection{Maxwell equations}
The Faraday tensor may be written in terms of the anti-symmetrised derivative of a vector potential
\begin{equation}
\label{eq:Faraday}
  F_{\alpha\beta} = 2\nabla_{[\alpha} A_{\beta]}.
\end{equation}
In terms of the vector potential, the Maxwell equations, $\nabla_{\alpha} F^{\alpha\beta} = J^\beta$ are given by
\begin{equation}
\label{eq:MaxwellPotential}
  (\mathcal{E} A)^\beta \equiv 2 \nabla_\alpha \nabla^{[\alpha} A^{\beta]} = J^\beta.
\end{equation}

\subsection{Teukolsky equations}

The Maxwell scalars (i.e. the tetrad components of the Faraday tensor) may be written in GHP form as
\begin{subequations}
\begin{align}
  \phi_0 & \equiv F_{lm} = \mathcal{T}_0 A, \label{eq:phi0} \\
  \phi_1 & \equiv \tfrac12\big(F_{ln}- F_{m\mb}\big) = \mathcal{T}_1 A, \label{eq:phi1} \\
  \phi_2 & \equiv F_{\mb n} \equiv \mathcal{T}_2 A \label{eq:phi2}
\end{align}
\end{subequations}
where the operators $\mathcal{T}_I$ are given by
\begin{subequations}
\begin{align}
  \mathcal{T}_0 A &= - (\edth-\bar{\tau}') A_l + (\th - \bar{\rho})A_m,\label{eq:T0} \\
  \mathcal{T}_1 A &= \frac12 \Big[- (\th' + \rho' - \bar{\rho}')A_l + (\th +\rho - \bar{\rho})A_n \nonumber \\
    & \qquad + (\edth' +\tau' -\bar{\tau})A_m - (\edth + \tau - \bar{\tau}')A_\mb \Big]  ,\label{eq:T1} \\
  \mathcal{T}_2 A &= (\edth'-\bar{\tau}) A_n -(\th' - \bar{\rho}')A_{\bar{m}}. \label{eq:T2}
\end{align}
\end{subequations}
We will later also need the adjoints of these, which are given by
\begin{subequations}
\begin{align}
  (\mathcal{T}^\dag_0 \Phi)^\alpha &= l^\alpha (\edth-\tau) \Phi - m^\alpha (\th - \rho) \Phi,\label{eq:T0dag} \\
  (\mathcal{T}^\dag_1 \Phi)^\alpha &= \frac12 \Big[l^\alpha (\th' - 2 \rho') \Phi -n^\alpha (\th - 2\rho) \Phi \nonumber \\
    & \qquad - m^\alpha (\edth' - 2\tau') \Phi + \mb^\alpha (\edth - 2\tau) \Phi \Big]  ,\label{eq:T1dag} \\
  (\mathcal{T}^\dag_2 \Phi)^\alpha &= - n^\alpha(\edth'-\tau') \Phi + \mb^\alpha (\th' - \rho')\Phi. \label{eq:T2dag}
\end{align}
\end{subequations}

The Maxwell scalars have types inherited from the tetrad vectors that appear in their definition,
\begin{equation}
  \phi_0 : \{2,0\}, \quad \phi_1 : \{0,0\}, \quad \phi_2 : \{-2,0\}.
\end{equation}
The scalars $\phi_0$ and $\phi_2$ satisfy the Teukolsky equations,
\begin{alignat}{4}
  \mathcal{O} \phi_0 &= \mathcal{S}_0 J,& \qquad \mathcal{O}' \phi_2 &= \mathcal{S}_2 J,
\end{alignat}
where
\begin{align}
  \mathcal{O} &\equiv \big(\th - 2 \,s\, \rho - \bar{\rho}\big)\big(\th'-\rho'\big) - \big(\edth - 2\, s \,\tau - \bar{\tau}'\big)\big(\edth' - \tau'\big) \nonumber \\
  &\qquad + \tfrac12 \big[\big(6 s-2\big)-4s^2\big] \psi_2
\end{align}
is the spin-weight $s$ Teukolsky operator.\footnote{Some authors (e.g. \cite{Wald:1978vm}) define $\mathcal{O}$ to be the operator with $s=+1$. Then, the operator for the negative $s$ field is its adjoint.} The middle scalar satisfies the Fackerell-Ipser equation \cite{Fackerell:1972hg},
\begin{equation}
    \mathcal{O}_1 \phi_1 = \mathcal{S}_1 J
\end{equation}
where
\begin{equation}
 \mathcal{O}_1 \equiv \big(\th - \rho - \bar{\rho}\big)\big(\th'-2\rho'\big) - \big(\edth - \tau - \bar{\tau}'\big)\big(\edth' - 2 \tau'\big)
\end{equation}
is the Fackerell-Ipser operator. The decoupling operators\footnote{Note that the middle operator can be decomposed into two parts, $\mathcal{S}_1 = \frac12(q - q')$, where $q = \tfrac12 \big[(\th' - \rho' - \bar{\rho}') J_l - (\edth'-\tau'-\bar{\tau}) J_m \big]$. Both $q$ and $q'$ lead to the same decoupled equations as $\mathcal{S}_1$ and satisfy the same Wald-type identity.}
\begin{subequations}
\begin{align}
  \mathcal{S}_0 J &= \tfrac12 \big[(\edth-2\tau-\bar{\tau}') J_l-(\th - 2\rho - \bar{\rho}) J_m\big], \label{eq:S0}\\
  \mathcal{S}_1 J &= \tfrac14 \big[(\th' - \rho' - \bar{\rho}') J_l - (\th - \rho - \bar{\rho}) J_n \nonumber \\ 
    & \qquad - (\edth'-\tau'-\bar{\tau}) J_m + (\edth-\tau-\bar{\tau}') J_\mb \big], \label{eq:S1} \\
  \mathcal{S}_2 J &= \tfrac12 \big[-(\edth'-2\tau' - \bar{\tau}) J_n + (\th' - 2 \rho' - \bar{\rho}') J_\mb\big]. \label{eq:S2} 
\end{align}
\end{subequations}
give the sources in terms of the 4-current. We will later also need the adjoints of these, which are given by
\begin{subequations}
\begin{align}
  \label{eq:S0dag}
  (\mathcal{S}^\dag_0 \Phi)_\alpha &= \frac12 \big[- l_\alpha (\edth+\tau) + m_\alpha (\th + \rho) \big]\Phi , \\
  (\mathcal{S}^\dag_1 \Phi)_\alpha &= \tfrac14 \big[- l^\alpha \th' + n^\alpha \th + m^\alpha \edth' - \mb^\alpha \edth \big]\Phi, \label{eq:S1dag} \\
  \label{eq:S2dag}
  (\mathcal{S}^\dag_2 \Phi)_\alpha &= \frac12 \big[n_\mu (\edth'+ \tau') - \mb_\mu (\th' + \rho') \big]\Phi .
\end{align}
\end{subequations}

It is worth noting that $\mathcal{O}' \phi_2 = \zeta^{-2} \mathcal{O} \zeta^2 \phi_2$ and $\mathcal{O} \phi_0 = \zeta^{-2} \mathcal{O}' \zeta^{2} \phi_0$, that $2 \mathcal{S}_I = -\zeta^{-2} \mathcal{T}_I \zeta^2$, and that $2 \mathcal{S}^\dag_I = -\zeta^{2} \mathcal{T}^\dag_I \zeta^{-2}$. Additionally, the Teukolsky and vector wave operators can be written in the simple factored forms
\begin{subequations}
\begin{align}
    \mathcal{O} \phi_0 &= -2 \mathcal{S}_0 \mathcal{T}^\dag_{2} \phi_0, \\
    \mathcal{O}_1 \phi_1 &= 4 \mathcal{S}_1 \mathcal{T}^\dag_{1} \phi_1, \\
    \mathcal{O}' \phi_2 &= -2 \mathcal{S}_2 \mathcal{T}^\dag_{0} \phi_2,
\end{align}
\end{subequations}
and
\begin{equation}
  \mathcal{E} = - 2 \mathcal{T}_0^\dag \mathcal{T}_2 + 4 \mathcal{T}_1^\dag \mathcal{T}_1 -2 \mathcal{T}_2^\dag \mathcal{T}_0.
\end{equation}
Using the fact that $\mathcal{S}_0 \mathcal{T}^\dag_0$, $\mathcal{S}_0 \mathcal{T}^\dag_1$, $\mathcal{S}_1 \mathcal{T}^\dag_0$, $\mathcal{S}_1 \mathcal{T}^\dag_2$, $\mathcal{S}_2 \mathcal{T}^\dag_2$ and $\mathcal{S}_2 \mathcal{T}^\dag_1$ all vanish as a consequence of the GHP commutators, it is immediately clear that Wald's operator identities hold \cite{Wald:1978vm},
\begin{equation}
    \mathcal{S}_0 \mathcal{E} = \mathcal{O} \mathcal{T}_0, \quad
    \mathcal{S}_1 \mathcal{E} = \mathcal{O}_1 \mathcal{T}_1, \quad
    \mathcal{S}_2 \mathcal{E} = \mathcal{O}' \mathcal{T}_2.
\end{equation}

In vacuum Kerr-NUT spacetimes \cite{Newman:1963yy}, the Teukolsky operator may be written in manifestly separable form by rewriting it in terms of the commuting symmetry operators \cite{Aksteiner:2014zyp}
\begin{equation}
  \mathscr{R} \equiv \zeta \bar{\zeta} (\th - \rho - \bar{\rho})(\th' - 2 b \rho') + \frac{2b-1}{2} (\zeta + \bar{\zeta}) \mathcal{\pounds}_\xi,
\end{equation}
and
\begin{equation}
  \mathscr{S} \equiv \zeta \bar{\zeta} (\edth - \tau - \bar{\tau}')(\edth' - 2 s \tau') + \frac{2s-1}{2} (\zeta - \bar{\zeta}) \mathcal{\pounds}_\xi.
\end{equation}
Then, the Teukolsky operator is given by
\begin{equation}
   \zeta \bar{\zeta} \mathcal{O} = \mathscr{R} - \mathscr{S}.
\end{equation}
The symmetry operators satisfy the commutation relations $\big[\mathscr{R}, \mathscr{S}\big] = 0$ when acting on a type $\{p,0\}$ object. We will see later that when written as a coordinate expression in Boyer-Lindquist coordinates in Kerr spacetime the operators $\mathscr{R}$ and $\mathscr{S}$ reduce to the radial Teukolsky and spin-weighted spheroidal operators (with a common eigenvalue).

\subsection{Teukolsky-Starobinsky identities}

In the homogeneous case, $J^\alpha = 0$, the Teukolsky-Starobinsky identities that relate $\phi_0$ to $\phi_2$ are given in GHP form by
\begin{subequations}
\label{eq:TS-EM}
\begin{align}
\label{eq:TS-E1}
  \th^2 \zeta^2 \phi_2 &= \edth'^2 \zeta^2 \phi_0, \\
\label{eq:TS-EM2}
  \th'^2 \zeta^2 \phi_0 &= \edth^2 \zeta^2 \phi_2, \\
\label{eq:TS-EM3}
  [\th' \edth' + \bar{\tau} \th'] \zeta^2 \phi_0 &= [\th \edth + \bar{\tau}' \th] \zeta^2\phi_2.
\end{align}
\end{subequations}
From these, we can also derive fourth-order Teukolsky-Starobinsky identities,
\begin{subequations}
\begin{align}
\label{eq:TS-EM4}
  \th^2 \bar{\zeta}^2 \th'^2 \zeta^2 \phi_0 &= \edth^2 \bar{\zeta}^2 \edth'^2 \zeta^2 \phi_0, \\
\label{eq:TS-EM5}
  \th'^2 \bar{\zeta}^2 \th^2 \zeta^2 \phi_2 &= \edth'^2 \bar{\zeta}^2 \edth^2 \zeta^2 \phi_2.
\end{align}
\end{subequations}
This latter form can be rewritten in terms of the symmetry operators,
\begin{subequations}
\begin{align}
  \th^2 \bar{\zeta}^2 \th'^2 \zeta^2 \phi_0 &= \big[\mathscr{R}^2 + \mathcal{\pounds}_\eta \mathcal{\pounds}_\xi \big] \phi_0, \\
  \edth^2 \bar{\zeta}^2 \edth'^2 \zeta^2 \phi_0 &= \big[\mathscr{S}^2 + \mathcal{\pounds}_\eta \mathcal{\pounds}_\xi \big] \phi_0, \\
  \th'^2 \bar{\zeta}^2 \th^2 \zeta^2 \phi_2 &= \big[\mathscr{R}'^2 + \mathcal{\pounds}_\eta \mathcal{\pounds}_\xi \big]\phi_2, \\
  \edth'^2 \bar{\zeta}^2 \edth^2 \zeta^2 \phi_2 &= \big[\mathscr{S}'^2 + \mathcal{\pounds}_\eta \mathcal{\pounds}_\xi \big] \phi_2.
\end{align}
\end{subequations}

\subsection{Reconstruction of a vector potential in radiation gauge}

In the ingoing radiation gauge (IRG), the vector potential may be reconstructed by applying a first-order differential operator to a type $\{-2,0\}$ scalar (i.e. the same type as $\phi_2$), $\Phi^{\rm IRG}$, called the Hertz potential. In terms of the Hertz potential, the IRG vector potential is given explicitly by
\begin{align}
  \label{eq:AIRG}
  A_\alpha^{\rm IRG} = 2 \Re [(\mathcal{S}^\dag_0 \Phi^{\rm IRG})_\alpha].
\end{align}
The IRG Hertz potential satisfies $\mathcal{O} \Phi^{\rm IRG} = \eta^{\rm IRG}$, where $\eta^{\rm IRG}$ satisfies $\Re (\mathcal{T}_0^\dag \eta^{\rm IRG})_\alpha = J_\alpha$. In other words, $\Phi^{\rm IRG}$ is a solution of the equation satisfied by $\zeta^2 \phi_2$ (equivalently, the adjoint of the equation satisfied by $\phi_0$), but with a different source.

The IRG vector potential manifestly satisfies the gauge condition $A_l = 0$ and it necessarily requires that $(\mathcal{E}A^{\rm IRG})_l = 0$.
Computing the Maxwell scalars from it, we find
\begin{subequations}
    \label{eq:phi-IRG}
\begin{align}
  \phi_0 &= \frac12 \th^2 \overline{\Phi^{\rm IRG}}, \label{eq:phi0-IRG}\\
  \phi_1 &= \frac12 [\th \edth' + \tau' \th] \overline{\Phi^{\rm IRG}}, \label{eq:phi1-IRG}\\
  \phi_2 &= \frac12 \edth'^2 \overline{\Phi^{\rm IRG}} - \frac12 \eta^{\rm IRG}. \label{eq:phi2-IRG}
\end{align}
\end{subequations}
The IRG Hertz potential may therefore be obtained either by solving the sourced (adjoint) Teukolsky equation or by solving either of the second-order equations sourced by the Maxwell scalars. The equations involving $\phi_0$ and $\phi_2$ are often referred to as the ``radial'' and ``angular'' inversion equations, respectively. Acting on the Maxwell scalars with the Teukolsky operator and commuting operators, we find
\begin{subequations}
\begin{align}
  \mathcal{O} \phi_0 &= \frac12 \Big(\th-\rho-\bar{\rho}\Big)^2 \overline{\eta^{\rm IRG}} ,\label{eq:Ophi0-IRG} \\
  \mathcal{O}_1 \phi_1 &= \frac12 \Big(\th-\rho-\bar{\rho}\Big)\Big(\edth' - \bar{\tau}\Big) \overline{\eta^{\rm IRG}} ,\label{eq:Ophi1-IRG} \\
  \mathcal{O}' \phi_2 &= \frac12 \Big(\edth'-\tau'-\bar{\tau}\Big)^2 \overline{\eta^{\rm IRG}} - \frac12 \mathcal{O}' \eta^{\rm IRG}. \label{eq:Ophi2-IRG}
\end{align}
\end{subequations}
Thus, the Maxwell scalars satisfy the homogeneous Teukolsky equation if $\mathcal{O} \Phi^{\rm IRG} =0 = \overline{\mathcal{O} \Phi^{\rm IRG}}$.

Similarly, in the outgoing radiation gauge (ORG) the vector potential is given by the prime of the IRG vector potential,
\begin{align}
  \label{eq:AORG}
  A_\alpha^{\rm ORG} = 2 \Re [(\mathcal{S}^\dag_2 \Phi^{\rm ORG})_\alpha]
\end{align}
where the ORG Hertz potential, $\Phi^{\rm ORG}$, is of type $\{2,0\}$ (i.e. the same as $\phi_0$). The ORG Hertz potential satisfies $\mathcal{O}' \Phi^{\rm ORG} = \eta^{\rm ORG}$, where $\eta^{\rm ORG}$ satisfies $\Re (\mathcal{T}_2^\dag \eta^{\rm ORG})_\alpha = J_\alpha$. In other words, $\Phi^{\rm ORG}$ is a solution of the equation satisfied by $\zeta^{2} \phi_0$ (equivalently, the adjoint of the equation satisfied by $\phi_2$), but with a different source.

The ORG vector potential manifestly satisfies the gauge condition $A_n = 0$ and it necessarily requires that $(\mathcal{E}A^{\rm IRG})_n = 0$.
Computing the Maxwell scalars from it, we find
\begin{subequations}
  \label{eq:phi-ORG}
\begin{align}
  \phi_0 &=  - \frac12 \edth^2 \overline{\Phi^{\rm ORG}} + \frac12 \eta^{\rm ORG}, \label{eq:phi0-ORG}\\
  \phi_1 &= \frac12 [\th' \edth + \tau \th'] \overline{\Phi^{\rm ORG}}, \label{eq:phi1-ORG}\\
  \phi_2 &= - \frac12 \th'^2 \overline{\Phi^{\rm ORG}}. \label{eq:phi2-ORG}
\end{align}
\end{subequations}
The ORG Hertz potential may therefore be obtained either by solving the sourced (adjoint) Teukolsky equation or by solving either of the second-order equations sourced by the Maxwell scalars. The equations involving $\phi_0$ and $\phi_2$ are often referred to as the ``angular'' and ``radial'' inversion equations, respectively. Acting on the Maxwell scalars with the Teukolsky operator and commuting operators, we find
\begin{subequations}
\begin{align}
  \mathcal{O} \phi_0 & = - \frac12 \Big(\edth-\tau-\bar{\tau}'\Big)^2 \overline{\eta^{\rm ORG}} + \frac12 \mathcal{O} \eta^{\rm ORG}, \label{eq:Ophi0-ORG} \\
  \mathcal{O}_1 \phi_1 &= \frac12 \Big(\th'-\rho'-\bar{\rho}'\Big)\Big(\edth - \bar{\tau}'\Big) \overline{\eta^{\rm ORG}},\label{eq:Ophi1-ORG} \\
  \mathcal{O}' \phi_2 &=
   -\frac12 \Big(\th'-\rho'-\bar{\rho}'\Big)^2 \overline{\eta^{\rm ORG}}. \label{eq:Ophi2-ORG}
\end{align}
\end{subequations}
Thus, the Maxwell scalars satisfy the homogenous Teukolsky equation if $\mathcal{O}' \Phi^{\rm ORG} =0 = \overline{\mathcal{O}' \Phi^{\rm ORG}}$.

In the vacuum case, the IRG and ORG Hertz potentials are not independent, but are related by Teukolsky-Starobinsky identities, much like the  Maxwell scalars themselves. To obtain identities for the Hertz potentials we use the fact that the Maxwell scalars are gauge invariant to equate Eqs.~\eqref{eq:phi-IRG} and \eqref{eq:phi-ORG} to give
\begin{subequations}
\begin{align}
  \frac12 \th^2 \overline{\Phi^{\rm IRG}} &= - \frac12 \edth^2 \overline{\Phi^{\rm ORG}} + \frac12 \eta^{\rm ORG}, \\
  \frac12 [\th \edth' + \tau' \th] \overline{\Phi^{\rm IRG}} &=  \frac12 [\th' \edth + \tau \th'] \overline{\Phi^{\rm ORG}}, \\
  \frac12 \edth'^2 \overline{\Phi^{\rm IRG}} - \frac12 \eta^{\rm IRG} &=  - \frac12 \th'^2 \overline{\Phi^{\rm ORG}}.
\end{align}
\end{subequations}

Finally, it is also possible to reconstruct a vector potential from a scalar $\Phi_1$ of type $\{0,0\}$ (i.e. the same type as the middle scalar $\phi_1$). Depending on whether one uses $\mathcal{S}_1^\dag$, $q^\dag$ or $q'^\dag$ the reconstructed vector potential is in radiation gauge (for $q^\dag$ and $q'^\dag$) or not (for $\mathcal{S}_1^\dag$). All three possibilities are also not in Lorenz gauge and do not necessarily require that any components of $(\mathcal{E}A^{\rm IRG})^\alpha$ are zero. In all three cases, the middle Hertz potential satisfies $\zeta^2 \mathcal{O}_1 \zeta^{-2} \Phi_1 = \eta_1$, where $\eta_1$ satisfies $\Re (\mathcal{T}_1^\dag \eta_1)_\alpha = J_\alpha$. In other words, $\Phi_1$ is a solution of the equation satisfied by $\zeta^{2} \phi_1$ (equivalently, the adjoint of the equation satisfied by $\phi_1$), but with a different source.

Computing the Maxwell scalars in this case we find
\begin{subequations}
\begin{align}
  \phi_0 &=  \frac12 (\edth - \bar{\tau}')\th \overline{\Phi_1}, \label{eq:phi0-NRG}\\
  \phi_1 &= \frac14 \big[(\th' + \rho' - \bar{\rho}')\th + (\edth + \tau - \bar{\tau}')\edth' \big] \overline{\Phi_1} + \frac14 \eta_1, \label{eq:phi1-NRG}\\
  \phi_2 &= \frac12 (\edth' - \bar{\tau})\th' \overline{\Phi_1}. \label{eq:phi2-NRG}
\end{align}
\end{subequations}
The middle Hertz potential may therefore be obtained either by solving the sourced (adjoint) Teukolsky equation or by solving either of the second-order equations sourced by the Maxwell scalars. Acting on the Maxwell scalars with the Teukolsky operator and commuting operators, we find
\begin{subequations}
\begin{align}
  \mathcal{O} \phi_0 & = \frac12 \big[(\edth-\tau-2\bar{\tau}')(\th - \rho - \bar{\rho}) - \bar{\tau}' \bar{\rho}\big] \overline{\eta_1}, \label{eq:Ophi0-NRG} \\
  \mathcal{O}_1 \phi_1 &= \frac14 \mathcal{O}_1 \eta_1 + \frac14 \big[(\th - \rho - \bar{\rho} ) (\th' - 2 \bar{\rho}') \nonumber  \\
    & \qquad \qquad \qquad + (\edth - \tau - \bar{\tau}' ) (\edth' - 2 \bar{\tau}) \big] \overline{\eta_1},\label{eq:Ophi1-NRG} \\
  \mathcal{O}' \phi_2 & = \frac12 \big[(\edth'-\tau'-2\bar{\tau})(\th' - \rho' - \bar{\rho}') - \bar{\tau} \bar{\rho}'\big] \overline{\eta_1}. \label{eq:Ophi2-NRG}
\end{align}
\end{subequations}
Thus, the Maxwell scalars satisfy the homogenous homogeneous Teukolsky equation if $\zeta^2 \mathcal{O}_1 \zeta^{-2} \Phi_1 =0 = \overline{\zeta^2 \mathcal{O}_1 \zeta^{-2} \Phi_1}$, i.e. $\Phi_1$ is a homogenous solution of the equation satisfied by $\zeta^{2} \phi_1$. Unfortunately, this third reconstruction approach is less useful than the other two as the Fackerell-Ipser equation is not known to be separable.

The fact that these potentials are solutions of the homogeneous Maxwell equations was succinctly summarised by Wald \cite{Wald:1978vm} using the method of adjoints: since the operators satisfy the identity $\mathcal{S} \mathcal{E} = \mathcal{O} \mathcal{T}$, by taking the adjoint and using the fact that $\mathcal{E}$ is self-adjoint we find that $\mathcal{E} \mathcal{S}^\dag = \mathcal{T}^\dag \mathcal{O}^\dag$ so we have a homogeneous solution of the Maxwell equations provided the Hertz potential satisfies the (adjoint) homogeneous Teukolsky equation.

\section{Lorenz gauge Hertz potentials}
\label{sec:MaxwellHertz}

In Lorenz gauge the vector potential satisfies the vector wave equation
\begin{equation}
\label{eq:VectorWave}
  (\mathcal{L}A)^\alpha \equiv \Box A^\alpha - R^\alpha{}_\beta A^\beta = J^\alpha,
\end{equation}
along with the Lorenz gauge condition
\begin{equation}
\label{eq:LorenzGauge}
  \nabla^\alpha A_\alpha=0.
\end{equation}
In a vacuum type-D spacetime, these may be written in GHP form as
\begin{subequations}
\begin{align}
\label{eq:VectorWaveGHP}
  (\mathcal{L}A)_l &= 
    2 \big[(\edth - \bar{\tau}')(\edth'-\tau')- (\th - \bar{\rho})(\th' - \rho') + \bar{\rho}\bar{\rho}' \big] A_l \nonumber \\
    & \quad + 2 \rho \bar{\rho} A_n + 2 \big[\bar{\rho} \edth' - \bar{\tau} \th \big] A_m + 2 \big[\rho \edth - \tau \th \big] A_\mb,
    \\
  (\mathcal{L}A)_m &= 
    2 \big[(\edth - \bar{\tau}')(\edth'-\tau')- (\th - \bar{\rho})(\th' - \rho') - \bar{\tau}\bar{\tau}'\big] A_m \nonumber \\
    & \quad - 2 \tau \bar{\tau}' A_\mb + 2 \big[\bar{\rho}' \edth - \bar{\tau}' \th' \big] A_l + 2 \big[\rho \edth - \tau \th \big] A_n,
    \\
  (\mathcal{L}A)_n &= \overline{(\mathcal{L}A)_{l}'},\\
  (\mathcal{L}A)_\mb &= \overline{(\mathcal{L}A)_{m}}
\end{align}
\end{subequations}
and
\begin{align}
\label{eq:LorenzGaugeGHP}
    \nabla^\alpha A_\alpha &= 
    (\edth' - \tau' - \bar{\tau}) A_m
    + (\edth - \tau - \bar{\tau}' A_\mb \nonumber \\
    & \quad - (\th'- \rho' - \bar{\rho}') A_l
    - (\th - \rho - \bar{\rho}) A_n = 0.
\end{align}

\subsection{Hertz potential derived from a two-form}
\label{sec:LorenzHertzTwoForm}

In order to identify a Hertz potential for the Lorenz gauge vector potential, we start with a real tensor with the same symmetries as the Faraday tensor (i.e. a two-form): $H_{\alpha\beta} = H_{[\alpha\beta]}$. This can be decomposed onto a null tetrad,
\begin{align}
  &H_{\alpha\beta} = 2 \Big[
      (\Phi^{\mathcal{L}1}_1+\bar{\Phi}^{\mathcal{L}1}_1) n_{[\alpha} l_{\beta]}  
    + (\Phi^{\mathcal{L}1}_1-\bar{\Phi}^{\mathcal{L}1}_1) m_{[\alpha} \bar{m}_{\beta]} \nonumber \\
   &\,\, + \Phi^{\mathcal{L}1}_0 \bar{m}_{[\alpha} n_{\beta]}
        + \bar{\Phi}^{\mathcal{L}1}_0 m_{[\alpha} n_{\beta]}
        + \Phi^{\mathcal{L}1}_2 l_{[\alpha} m_{\beta]} 
        + \bar{\Phi}^{\mathcal{L}1}_2 l_{[\alpha} \bar{m}_{\beta]}
     \Big],
\end{align}
where the GHP type of the complex scalars is the same as that of the Maxwell scalars:
\begin{align}
  &\Phi^{\mathcal{L}1}_0 : \{2,0\}, \quad \Phi^{\mathcal{L}1}_1 : \{0,0\}, \quad \Phi^{\mathcal{L}1}_2 : \{-2,0\}, \nonumber \\
  &\bar{\Phi}^{\mathcal{L}1}_0 : \{0,2\}, \quad \bar{\Phi}^{\mathcal{L}1}_1 : \{0,0\}, \quad \bar{\Phi}^{\mathcal{L}1}_2 : \{0,-2\}.
\end{align}
Note that under the GHP prime operation we have the identification $\Phi_0^{\mathcal{L}1} = -(\Phi^{\mathcal{L}1}_2)'$, where the minus sign arises from the fact that $H_{\alpha\beta}$ is anti-symmetric.
We will allow this tensor to only have maximum spin-weight components, i.e. $\Phi^{\mathcal{L}1}_1 = 0 = \bar{\Phi}^{\mathcal{L}1}_1$ and decompose it into self-dual and anti-self-dual parts, $H_{\alpha \beta} = \mathcal{H}_{\alpha\beta} + \bar{\mathcal{H}}_{\alpha\beta}$, where
\begin{subequations}
\begin{align}
    \mathcal{H}_{\alpha\beta} &= \frac12(H_{\alpha\beta} - i\,{}^\ast H_{\alpha\beta}),\\
    \bar{\mathcal{H}}_{\alpha\beta} &= \frac12(H_{\alpha\beta} + i\,{}^\ast H_{\alpha\beta}),
\end{align}
\end{subequations}
where ${}^\ast H_{\alpha\beta} = \frac{1}{2} \epsilon_{\alpha\beta}{}^{\gamma\delta} H_{\gamma\delta}$ is the Hodge dual of $H_{\alpha\beta}$, and where the (anti-)self-dual property means that ${}^\ast \mathcal{H}_{\alpha\beta} = i \mathcal{H}_{\alpha\beta}$ and ${}^\ast \bar{\mathcal{H}}_{\alpha\beta} = -i \bar{\mathcal{H}}_{\alpha\beta}$. The self-dual part depends only on $\Phi^{\mathcal{L}1}_0$ and $\Phi^{\mathcal{L}1}_2$ and the anti-self-dual part depends only on $\bar{\Phi}^{\mathcal{L}1}_0$ and $\bar{\Phi}^{\mathcal{L}1}_2$,
\begin{subequations}
\begin{align}
  \mathcal{H}_{\alpha\beta} &= 2\Big[\Phi^{\mathcal{L}1}_0 \mb_{[\alpha} n_{\beta]} + \Phi^{\mathcal{L}1}_2 l_{[\alpha} m_{\beta]}\Big], \\
  \bar{\mathcal{H}}_{\alpha\beta} &= 2\Big[\bar{\Phi}^{\mathcal{L}1}_0 m_{[\alpha} n_{\beta]} + \bar{\Phi}^{\mathcal{L}1}_2 l_{[\alpha} \bar{m}_{\beta]}\Big].
\end{align}
\end{subequations}
Now construct a complex vector\footnote{It is, of course, possible to obtain a real vector from this complex potential and its complex conjugate.}
by taking the divergence of $\zeta\mathcal{H}_{ab}$
\begin{equation}
\label{eq:AL1}
  A^{\mathcal{L}1}_{\alpha} = \nabla^\beta (\zeta \mathcal{H}_{\beta\alpha}).
\end{equation}
This vector has tetrad components
\begin{subequations}
\label{eq:AL1GHP}
\begin{align}
  A^{\mathcal{L}1}_{l}   &=  - \zeta (\edth' - 2 \tau') \Phi^{\mathcal{L}1}_0, \\
  A^{\mathcal{L}1}_{n}   &=  \zeta(\edth - 2 \tau) \Phi^{\mathcal{L}1}_2, \\
  A^{\mathcal{L}1}_{m}   &=  - \zeta (\th' - 2 \rho') \Phi^{\mathcal{L}1}_0, \\
  A^{\mathcal{L}1}_{\mb} &=  \zeta (\th - 2 \rho) \Phi^{\mathcal{L}1}_2.
\end{align}
\end{subequations}
It is straightforward to check that this vector satisfies the Lorenz gauge condition,
\begin{equation}
  \nabla^\alpha A_\alpha^{\mathcal{L}1} = 0,
\end{equation}
as a consequence of the GHP equations and the GHP commutators without assuming anything further about the scalars.

If we now compute the Maxwell scalars from this vector potential, we find
\begin{subequations}
\begin{align}
\label{eq:psi0Phi0} 
  \phi_0^{\mathcal{L}1} &= (-\zeta\mathcal{O} + \mathcal{\pounds}_\xi ) \Phi^{\mathcal{L}1}_0, \\
\label{eq:phi1Phi02} 
  \phi_1^{\mathcal{L}1} &= \zeta\big[ (\rho \edth - \tau \th) \Phi^{\mathcal{L}1}_2  + (\rho' \edth' - \tau' \th') \Phi^{\mathcal{L}1}_0\big], \\
\label{eq:phi2Phi2}
  \phi_2^{\mathcal{L}1} &= (-\zeta \mathcal{O}' - \mathcal{\pounds}_\xi) \Phi^{\mathcal{L}1}_2.
\end{align}
\end{subequations}
Acting on both sides with the appropriate Teukolsky operator and commuting the operators on the right hand side (recalling that $\mathcal{\pounds}_\xi$ commutes with everything), then we find
\begin{subequations}
\begin{align}
  \mathcal{O} \phi_0^{\mathcal{L}1} &= (-\mathcal{O}\zeta + \mathcal{\pounds}_\xi) \mathcal{O} \Phi^{\mathcal{L}1}_0,\\
  \mathcal{O}_1 \phi_1^{\mathcal{L}1} &=  \zeta\big[ (\rho \edth - \tau \th) \mathcal{O}'\Phi^{\mathcal{L}1}_2  + (\rho' \edth' - \tau' \th') \mathcal{O}\Phi^{\mathcal{L}1}_0\big] \nonumber \\ & \qquad
    - 2 \zeta^{-1}(\rho' \tau I^{\mathcal{L}1}_1 + \rho \tau' I^{\mathcal{L}1}_2) \nonumber \\ & \qquad
    + \big[(\zeta\bar{\zeta})^{-2} + 2 \zeta^{-1}(\rho \rho'+ \tau \tau')\big] I^{\mathcal{L}1}_3,\\
  \mathcal{O}' \phi_2^{\mathcal{L}1} &= (-\mathcal{O}'\zeta - \mathcal{\pounds}_\xi) \mathcal{O}' \Phi^{\mathcal{L}1}_2,
\end{align}
\end{subequations}
where
\begin{subequations}
\label{eq:TS-EM-L1}
\begin{align}
\label{eq:TS-EM1-L1}
  I^{\mathcal{L}1}_1 &\equiv \th^2 \zeta^2 \Phi^{\mathcal{L}1}_2 + \edth'^2 \zeta^2 \Phi^{\mathcal{L}1}_0, \\
\label{eq:TS-EM2-L1}
  I^{\mathcal{L}1}_2 &\equiv \th'^2 \zeta^2 \Phi^{\mathcal{L}1}_0 + \edth^2 \zeta^2 \Phi^{\mathcal{L}1}_2, \\
\label{eq:TS-EM3-L1}
  I^{\mathcal{L}1}_3 &\equiv [\th' \edth' + \bar{\tau} \th'] \zeta^2 \Phi^{\mathcal{L}1}_0 + [\th \edth + \bar{\tau}' \th] \zeta^2 \Phi^{\mathcal{L}1}_2.
\end{align}
\end{subequations}
closely resemble the Teukolsky-Starobinsky identities given in Eq.~\eqref{eq:TS-EM}.
We therefore find that the Maxwell scalars will satisfy the homogeneous Teukolsky equations if the Hertz potential scalars satisfy the Teukolsky equations, $\mathcal{O} \Phi^{\mathcal{L}1}_0 = 0 = \mathcal{O}' \Phi^{\mathcal{L}1}_2$, and are related by the Teukolsky-Starobinsky-like identities, $I^{\mathcal{L}1}_1 = 0$, $I^{\mathcal{L}1}_2 = 0$ and $I^{\mathcal{L}1}_3 = 0$. In that case, the relationship between the scalars and the two maximum-spin Maxwell scalars simplifies to
\begin{subequations}
\begin{align}
  \label{eq:phi0-Phi0-L1}
  \phi_{0}^{\mathcal{L}1} &= \mathcal{\pounds}_\xi \Phi_{0}^{\mathcal{L}1}, \\
  \label{eq:phi2-Phi2-L1}
  \phi_{2}^{\mathcal{L}1} &= - \mathcal{\pounds}_\xi \Phi_{2}^{\mathcal{L}1}.
\end{align}
\end{subequations}
Note that up to now we have only considered the contribution from the self-dual part of the Hertz potential, $\mathcal{H}_{\alpha\beta}$, and not from the anti-self-dual part $\bar{\mathcal{H}}_{\alpha\beta}$. However, the anti-self-dual Hertz potential does not contribute to the Maxwell scalars in the vacuum case, since
\begin{subequations}
\begin{align}
\label{eq:psi0Phi0b} 
  \phi_0^{\mathcal{L}1}\big(\overline{A^{\mathcal{L}1}}\big) &= \overline{\zeta^{-1} I^{\mathcal{L}1}_1} = 0, \\
\label{eq:psi1Phi0b} 
  \phi_1^{\mathcal{L}1}\big(\overline{A^{\mathcal{L}1}}\big) &= \overline{\zeta^{-1} I^{\mathcal{L}1}_3} = 0, \\
\label{eq:Psi2Phi2b}
  \phi_2^{\mathcal{L}1}\big(\overline{A^{\mathcal{L}1}}\big) &= \overline{\zeta^{-1} I^{\mathcal{L}1}_2} = 0.
\end{align}
\end{subequations}

Finally, substituting the vector potential into the Lorenz gauge field equations we find
\begin{subequations}
\label{eq:LorenzL1}
\begin{align}
    (\mathcal{L}A^{\mathcal{L}1})_l &= 2 \zeta (\edth' - 3 \tau') \mathcal{O} \Phi^{\mathcal{L}1}_0 + 2 \zeta^{-1} \rho I^{\mathcal{L}1}_3 - 2 \zeta^{-1} \tau I^{\mathcal{L}1}_1, \\
    (\mathcal{L}A^{\mathcal{L}1})_n &= -2 \zeta (\edth - 3 \tau) \mathcal{O}' \Phi^{\mathcal{L}1}_2 - 2 \zeta^{-1} \rho' I^{\mathcal{L}1}_3 + 2 \zeta^{-1} \tau' I^{\mathcal{L}1}_2, \\
    (\mathcal{L}A^{\mathcal{L}1})_m &= 2 \zeta (\th' - 3 \rho') \mathcal{O} \Phi^{\mathcal{L}1}_0 - 2 \zeta^{-1} \tau I^{\mathcal{L}1}_3 + 2 \zeta^{-1} \rho I^{\mathcal{L}1}_2, \\
    (\mathcal{L}A^{\mathcal{L}1})_\mb &= -2 \zeta (\th - 3 \rho) \mathcal{O}' \Phi^{\mathcal{L}1}_2 + 2 \zeta^{-1} \tau' I^{\mathcal{L}1}_3 - 2 \zeta^{-1} \rho' I^{\mathcal{L}1}_1.
\end{align}
\end{subequations}
This vector potential therefore satisfies the homogeneous Lorenz gauge field equations provided the scalars satisfy the homogeneous Teukolsky equation and the Teukolsky-Starobinsky-like identities. An alternative derivation of this final result is given in Appendix \ref{sec:LorenzHertzField}.

When decomposed into modes in Kerr spacetime, the potential $A^{\mathcal{L}1}_\alpha$ is the same as the one identified by Dolan \cite{Dolan:2019hcw}, and thus
so far we have merely reproduced his derivation in GHP form without relying on a mode decomposition.

\subsection{A second Lorenz-gauge Hertz potential}
\label{sec:LorenzHertzGHP}

We can arrive at the vector potential $A^{\mathcal{L}1}_\alpha$ by a different means. If we look for a vector potential that is constructed
from at most a first-order GHP operator acting on spin-weight $\pm 1$ scalars (i.e. scalars of type $\{2,0\}$, $\{0,2\}$, $\{-2,0\}$ and $\{0,-2\}$)
then the most general possibility is the complex potential
\begin{subequations}
\begin{align}
  \label{eq:ALAnsatzGHP}
  A_{l}   &=  (c_{l_1} \edth' + c_{l_2} \tau' + c_{l_3} \bar{\tau}) \Phi^{\mathcal{L}1}_0, \\
  A_{n}   &=  (c_{n_1} \edth  + c_{n_2} \tau + c_{n_3} \bar{\tau}') \Phi^{\mathcal{L}1}_2, \\
  A_{m}   &=  (c_{m_1} \th' + c_{m_2} \rho' + c_{m_3} \bar{\rho}') \Phi^{\mathcal{L}1}_0, \\
  A_{\mb} &=  (c_{\mb_1} \th + c_{\mb_2} \rho + c_{\mb_3} \bar{\rho}) \Phi^{\mathcal{L}1}_2,
\end{align}
\end{subequations}
and its complex conjugate. The coefficients here must be of GHP type $\{0,0\}$ (in particular, they may be constructed from numeric constants and functions of $\zeta$ and $\bar{\zeta}$). If we restrict to the case that the coefficients are linear functions of $\zeta$ and $\bar{\zeta}$ with arbitrary numeric constants, we find that the only possibility that satisfies the gauge condition and field equations is $A^{\mathcal{L}1}_a$.\footnote{Note that the gauge condition is satisfied unconditionally, but the field equations are only satisfied if the scalars satisfy the Teukolsky equation and Teukolsky-Starobinsky-like identities.}

This approach also allows us to identify a second Hertz potential by considering alternative forms for the coefficients. In particular, if we consider coefficients that are polynomials in $\zeta$ and $\bar{\zeta}$ we find a second complex potential,
\begin{subequations}
\begin{align}
  A^{\mathcal{L}2}_l &= \big[\tfrac{1}{2} \zeta (\zeta-\bar{\zeta})\edth' + \zeta \bar{\zeta} \tau' \big]\Phi^{\mathcal{L}2}_0,\\
  A^{\mathcal{L}2}_n &= \big[\tfrac{1}{2} \zeta(\zeta-\bar{\zeta}) \edth + \zeta \bar{\zeta} \tau \big]\Phi^{\mathcal{L}2}_2,\\
  A^{\mathcal{L}2}_m &= \big[\tfrac{1}{2} \zeta(\zeta+\bar{\zeta}) \th' - \zeta \bar{\zeta} \rho' \big]\Phi^{\mathcal{L}2}_0, \\
  A^{\mathcal{L}2}_\mb &= \big[\tfrac{1}{2} \zeta(\zeta+\bar{\zeta}) \th - \zeta \bar{\zeta} \rho \big]\Phi^{\mathcal{L}2}_2,
\end{align}
\end{subequations}
that yields the Lorenz gauge condition
\begin{equation}
  \nabla^\alpha A_\alpha^{\mathcal{L}2} = \bar{\zeta}\zeta^{-1} I^{\mathcal{L}2}_3.
\end{equation}
This second potential therefore satisfies the Lorenz gauge condition provided the scalars satisfy the same Teukolsky-Starobinsky-like identity as given in Eq.~\eqref{eq:TS-EM3-L1}. It is interesting to note that this second vector potential can also be written in tensor notation as
\begin{align}
  A^{\mathcal{L}2}_{\alpha} &= h^{\beta\gamma}\nabla_\beta (\zeta \mathcal{H}_{\alpha\gamma}),
\end{align}
where
\begin{equation}
  h_{\alpha \beta} = (\zeta + \bar{\zeta}) n_{[\alpha} l_{\beta]} - (\zeta - \bar{\zeta}) \mb_{[\alpha} m_{\beta]}
\end{equation}
is the conformal Killing-Yano tensor.

Computing the Maxwell scalars, we find
\begin{subequations}
\begin{align}
  \phi_0^{\mathcal{L}2} &= \frac12 \big( \zeta^2 \mathcal{O} + \mathscr{R} + \mathscr{S}\big) \Phi_0^{\mathcal{L}2},
  \\
  \phi_1^{\mathcal{L}2} &= \frac12 \zeta \bar{\zeta} \big[\rho(\edth+2\tau) - (\edth-\tau-\bar{\tau}')\th \big]\Phi_2^{\mathcal{L}2}
  \nonumber \\ & \quad
    - \frac12 \zeta \bar{\zeta} \big[\rho'(\edth'+2\tau') - (\edth'-\tau'-\bar{\tau})\th' \big] \Phi_0^{\mathcal{L}2},
  \\
  \phi_2^{\mathcal{L}2} &= -\frac12 \big( \zeta^2 \mathcal{O}' + \mathscr{R}' + \mathscr{S}' \big) \Phi_2^{\mathcal{L}2}.
\end{align}
\end{subequations}
Acting on both sides with the appropriate Teukolsky operator and commuting the operators on the right hand side, we then find
\begin{subequations}
\begin{align}
  \mathcal{O} \phi_0^{\mathcal{L}2} &= \frac12 \big[\mathcal{O}\zeta^2  + (\zeta\bar{\zeta})^{-1} (\mathscr{R}+\mathscr{S}) (\zeta\bar{\zeta} )\big] \mathcal{O}\Phi^{\mathcal{L}2}_0,\\
  \mathcal{O}_1 \phi_1^{\mathcal{L}2} &= \frac{\bar{\zeta}}{\zeta}(\rho \tau' I^{\mathcal{L}2}_2 - \rho' \tau I^{\mathcal{L}2}_1) \nonumber \\ &
  + \frac12 \zeta \bar{\zeta} \big[(\edth'-2\tau'-2 \bar{\tau})(\th'-2\rho'-\bar{\rho}')-2\rho'\tau' \big] \mathcal{O}\Phi^{\mathcal{L}2}_0  \nonumber \\ &
  -\frac12 \zeta \bar{\zeta} \big[(\edth-2\tau-2 \bar{\tau}')(\th-2\rho-\bar{\rho})-2\rho\tau \big] \mathcal{O}'\Phi^{\mathcal{L}2}_2,\\
  \mathcal{O}' \phi_2^{\mathcal{L}2} &= -\frac12 \big[\mathcal{O}'\zeta^2  + (\zeta\bar{\zeta})^{-1} (\mathscr{R}'+\mathscr{S}') (\zeta\bar{\zeta}) \big] \mathcal{O}'\Phi^{\mathcal{L}2}_2.
\end{align}
\end{subequations}
We therefore find that the Maxwell scalars will satisfy the homogeneous Teukolsky equations if the Hertz potential scalars satisfy the Teukolsky equations, $\mathcal{O} \Phi^{\mathcal{L}2}_0 = 0 = \mathcal{O}' \Phi^{\mathcal{L}2}_2$, and are related by the Teukolsky-Starobinsky-like identities, $I^{\mathcal{L}2}_1 = 0$, $I^{\mathcal{L}2}_2 = 0$ and $I^{\mathcal{L}2}_3 = 0$. Then, the relationship between our potentials and the two maximum-spin Maxwell scalars simplifies to
\begin{subequations}
\begin{align}
  \label{eq:phi0-Phi0-L2}
  \phi_{0}^{\mathcal{L}2} &= \frac12(\mathscr{R}+\mathscr{S}) \Phi_{0}^{\mathcal{L}2}, \\
  \label{eq:phi2-Phi2-L2}
  \phi_{2}^{\mathcal{L}2} &= -\frac12(\mathscr{R}'+\mathscr{S}') \Phi_{2}^{\mathcal{L}2}.
\end{align}
\end{subequations}
We recognise these as the radial and angular Teukolsky operators, meaning that when decomposed into modes in Kerr spacetime the potentials are simply the Maxwell scalars multiplied by the separation constant (eigenvalue). Note that in this case --- in contrast to the case with the first potential --- the complex conjugate potentials do contribute to the Maxwell scalars,
\begin{subequations}
\begin{align}
\label{eq:psi0Phi0bL2} 
  \phi_0^{\mathcal{L}2} \big(\overline{A^{\mathcal{L}2}}\big) &= \overline{- \edth'^2 \zeta^2 \Phi_0^{\mathcal{L}2} + \frac12 (1 + \bar{\zeta}\zeta^{-1}) I_1^{\mathcal{L}2}}, \\
\label{eq:psi1Phi0bL2} 
  \phi_1^{\mathcal{L}2} \big(\overline{A^{\mathcal{L}2}}\big) &= \overline{\frac12 \big[ (\th \edth + \bar{\tau}' \th) \zeta^2 \Phi_2^{\mathcal{L}2} - (\th' \edth' + \bar{\tau} \th') \zeta^2 \Phi_0^{\mathcal{L}2} \big]}, \\
\label{eq:Psi2Phi2bL2}
  \phi_2^{\mathcal{L}2} \big(\overline{A^{\mathcal{L}2}}\big) &= \overline{\edth^2 \zeta^2 \Phi_2^{\mathcal{L}2} - \frac12 (1 + \bar{\zeta}\zeta^{-1}) I_2^{\mathcal{L}2}}.
\end{align}
\end{subequations}

Finally, substituting the vector potential into the Lorenz gauge field equations we find
\begin{subequations}
\begin{align}
    (\mathcal{L}A^{\mathcal{L}2})_l &= - \zeta(\zeta - \bar{\zeta}) (\edth' - 3 \tau') \mathcal{O} \Phi^{\mathcal{L}2}_0 \nonumber \\ & \,
      + \zeta^{-1} (\zeta + \bar{\zeta})\rho I^{\mathcal{L}2}_3 - \zeta^{-1}(\zeta + \bar{\zeta}) \tau I^{\mathcal{L}2}_1, \\
    (\mathcal{L}A^{\mathcal{L}2})_n &= - \zeta(\zeta - \bar{\zeta}) (\edth - 3 \tau) \mathcal{O}' \Phi^{\mathcal{L}2}_2 \nonumber \\ & \,
      + \zeta^{-1} (\zeta + \bar{\zeta})\rho' I^{\mathcal{L}2}_3 - \zeta^{-1} (\zeta + \bar{\zeta})\tau' I^{\mathcal{L}2}_2, \\
    (\mathcal{L}A^{\mathcal{L}2})_m &= - \zeta (\zeta + \bar{\zeta})(\th' - 3 \rho') \mathcal{O} \Phi^{\mathcal{L}2}_0 \nonumber \\ & \,
      - \zeta^{-1}(\zeta - \bar{\zeta}) \tau I^{\mathcal{L}2}_3 + \zeta^{-1} (\zeta - \bar{\zeta}) \rho I^{\mathcal{L}2}_2, \\
    (\mathcal{L}A^{\mathcal{L}2})_\mb &= - \zeta (\zeta + \bar{\zeta})(\th - 3 \rho) \mathcal{O}' \Phi^{\mathcal{L}2}_2 \nonumber \\ & \,
      - \zeta^{-1} (\zeta - \bar{\zeta}) \tau' I^{\mathcal{L}2}_3 + \zeta^{-1} (\zeta - \bar{\zeta}) \rho' I^{\mathcal{L}2}_1.
\end{align}
\end{subequations}
This vector potential therefore satisfies the homogeneous Lorenz gauge field equations provided the scalars satisfy the homogeneous Teukolsky equation and the Teukolsky-Starobinsky-like identities. 

\subsection{Higher order Lorenz gauge potentials}
\label{sec:L3}

Clearly further potentials can be obtained trivially by taking linear combinations of the two previously identified potentials. Less trivially, we can also obtain new potentials by replacing the Hertz potential scalars with any of the symmetry
operators $\mathcal{\pounds}_\xi$, $\mathcal{\pounds}_\eta$, $\mathscr{R}$ and $\mathscr{S}$ applied to a new scalar of the same type. These new scalars will continue to satisfy the homogeneous Teukolsky equation and Teukolsky-Starobinsky-like identities since the symmetry operators commute with each other and with the rescaled Teukolsky operator $\zeta \bar{\zeta} \mathcal{O}$. 

These facts combined allow us to construct further,
higher order potentials by applying symmetry operators and taking linear combinations. For example, the potential
\begin{subequations}
\begin{align}
  \label{eq:AL3GHP}
  A^{\mathcal{L}3}_{l}   &=  - \zeta (\edth' - 2 \tau') \mathcal{\pounds}_\eta \Phi^{\mathcal{L}3}_0 \nonumber \\
   & \quad + \big[\tfrac12 \zeta(\zeta-\bar{\zeta}) \edth' + \zeta \bar{\zeta} \tau' \big] \mathscr{R} \Phi^{\mathcal{L}3}_0,
  \\
  A^{\mathcal{L}3}_{n}   &=  \zeta(\edth - 2 \tau) \mathcal{\pounds}_\eta \Phi^{\mathcal{L}3}_2  \nonumber \\
   & \quad + \big[\tfrac12 \zeta(\zeta-\bar{\zeta}) \edth + \zeta \bar{\zeta} \tau \big] \mathscr{R}' \Phi^{\mathcal{L}3}_2,
  \\
  A^{\mathcal{L}3}_{m}   &=  - \zeta (\th' - 2 \rho') \mathcal{\pounds}_\eta\Phi^{\mathcal{L}3}_0  \nonumber \\
   & \quad + \big[\tfrac12 \zeta(\zeta+\bar{\zeta}) \th' + \zeta \bar{\zeta} \rho' \big] \mathscr{R} \Phi^{\mathcal{L}3}_0,
  \\
  A^{\mathcal{L}3}_{\mb} &=  \zeta (\th - 2 \rho) \mathcal{\pounds}_\eta \Phi^{\mathcal{L}3}_2  \nonumber \\
   & \quad + \big[\tfrac12 \zeta(\zeta+\bar{\zeta}) \th + \zeta \bar{\zeta} \rho \big] \mathscr{R}' \Phi^{\mathcal{L}3}_2,
\end{align}
\end{subequations}
is also a homogeneous solution of the Lorenz gauge equations and satisfies the Lorenz gauge condition.
Note that this is constructed from a third-order operator acting on spin-weight $\pm 1$ scalars, so we are free to add terms involving a first order operator acting on the Teukolsky equation while still satisfying the Lorenz gauge equations.

Computing the corresponding maximum-spin Maxwell scalars we find
\begin{subequations}
\begin{align}
  \phi_0^{\mathcal{L}3} &= \frac12 \zeta \big[ (\zeta - \bar{\zeta}) (\zeta\bar{\zeta})^{-1} \mathscr{R} (\zeta\bar{\zeta})  - \zeta \mathcal{\pounds}_\eta \big] \mathcal{O}  \Phi_0^{\mathcal{L}3} \nonumber \\ & \qquad 
    + \th^2 \bar{\zeta}^2 \th'^2 \zeta^2 \Phi_0^{\mathcal{L}3}, \\
  \phi_2^{\mathcal{L}3} &= -\frac12 \zeta \big[ (\zeta - \bar{\zeta}) (\zeta\bar{\zeta})^{-1} \mathscr{R}' (\zeta\bar{\zeta})  - \zeta  \mathcal{\pounds}_\eta \big] \mathcal{O}' \Phi_2^{\mathcal{L}3} \nonumber \\ & \qquad 
    - \th'^2 \bar{\zeta}^2 \th^2 \zeta^2 \Phi_2^{\mathcal{L}3}.
\end{align}
\end{subequations}
In the homogeneous case, $\mathcal{O}  \Phi_0^{\mathcal{L}3} = 0 = \mathcal{O}' \Phi_2^{\mathcal{L}3}$ we recognise these as the fourth-order Teukolsky-Starobinsky operators, meaning that in modes the potentials are simply the Maxwell scalars divided by the Teukolsky-Starobinsky constant.

\section{Mode decomposed equations in Kerr spacetime}
\label{sec:Modes}

We now specialise the results of the previous sections to the case of mode-decomposed perturbations of Kerr spacetime. We work with the metric of Kerr spacetime in Boyer-Lindquist coordinates $(t,r,\theta,\varphi)$, which is given by
\begin{align}
\label{eq:KerrMetricBL}
ds^2 &= - \bigg[1-\frac{2Mr}{\Sigma}\bigg]dt^2 
			- \frac{4aMr\sin^2\theta}{\Sigma}dt\, d\varphi
			+ \frac{\Sigma}{\Delta}dr^2\nonumber \\ & \qquad
			+ \Sigma d\theta^2
			+ \bigg[\Delta+\frac{2Mr(r^2+a^2)} {\Sigma}\bigg] \sin^2\theta d\varphi^2,
\end{align}
where $\Sigma = r^2 + a^2 \cos^2 \theta$ and $\Delta = r^2 -2 M r +a^2$.

\subsection{Null tetrads in Kerr spacetime}
A null tetrad proposed by Kinnersley \cite{Kinnersley:1969zza} is a common choice when dealing with
perturbations of Kerr spacetime. Indeed, it formed a crucial part of Teukolsky's
separability result for perturbations of the Weyl tensor \cite{Teukolsky:1972my}. However, the
Kinnersley tetrad has two unfortunate features that make it less than ideal for elucidating the
symmetric structure of perturbations of Kerr spacetime: (i) it violates the
$\{t, \varphi\} \to \{-t,-\varphi\}$ symmetry; and (ii) it destroys a symmetry in $\{r, \theta\}$. An alternative tetrad proposed by Carter \cite{Carter:1987hk} does not suffer from either of these deficiencies. In terms of Boyer-Lindquist coordinates and using the convention of having $l^\alpha$ point outwards, Carter's tetrad has components
\begin{subequations}
\begin{align}
l^\alpha &= \frac{1}{\sqrt{2\Delta \Sigma}}\Big[r^2+a^2,\Delta,0,a\Big], \\
n^\alpha &= \frac{1}{\sqrt{2\Delta \Sigma}}\Big[r^2+a^2,-\Delta,0,a\Big], \\
m^\alpha &= \frac{1}{\sqrt{2\Sigma}}\Big[i a \sin \theta,0,1,\frac{i}{\sin \theta}\Big], \\
\bar{m}^\alpha &= \frac{1}{\sqrt{2\Sigma}}\Big[-i a \sin \theta,0,1,-\frac{i}{\sin \theta}\Big].
\end{align}
\end{subequations}
This corresponds to a simple rescaling of Kinnersley's tetrad (denoted with a K subscript):
$l^\alpha   = \sqrt{\Delta/2\Sigma}\,l^\alpha_\mathrm{K}$,
$n^\alpha   = \sqrt{2\Sigma/\Delta}\,n^\alpha_\mathrm{K}$,
$m^\alpha   = \bar{\zeta}/\sqrt{\Sigma}m^\alpha_\mathrm{K}$, $\mb^\alpha = \zeta/\sqrt{\Sigma} \mb^\alpha_\mathrm{K}$. This rescaling has the effect of causing the Carter tetrad to be more symmetric in that it transforms as $l \leftrightarrow - n$, $m \leftrightarrow \mb$ (note the minus sign means that this does not correspond to the GHP prime operation) under $\{t,\varphi\} \to \{-t, -\varphi\}$.

For the Carter tetrad the non-zero spin coefficients have a particularly symmetric form given by
\begin{subequations}
\begin{align}
 \rho = -\rho' &= -\frac{1}{\zeta} \sqrt{\frac{\Delta}{2 \Sigma}}, \\
 \tau = \tau' &= -\frac{i a \sin \theta}{\zeta\sqrt{2 \Sigma}} ,\\
 \beta  = \beta' &= -\frac{i}{\zeta} \frac{a+i r \cos \theta}{2\sin\theta\sqrt{2\Sigma}},\\
 \epsilon = - \epsilon' &= \frac{M r - a^2 - i a (r-M) \cos \theta}{2\zeta\sqrt{2 \Sigma \Delta}},
\end{align}
\end{subequations}
where in Kerr spacetime the Killing spinor coefficient is given by $\zeta = r-i a \cos \theta$.
The commuting GHP operators have the same form in both the Carter and Kinnersley tetrads, and are given by
\begin{subequations}
\begin{align}
  \mathcal{\pounds}_{\xi} &= \partial_t,\\
  \mathcal{\pounds}_{\eta} &= a^2 \partial_t + a \partial_\varphi.
\end{align}
\end{subequations}

Although the results in the following sections are given for the Carter tetrad, it is important to note that none of the fundamental conclusions
depend on this choice, and similar results follow when using, e.g., the Kinnersley tetrad.

\begin{widetext}
\subsection{Lorenz gauge vector potentials}

The tetrad components of the Lorenz gauge vector potentials have Boyer-Lindquist coordinate expressions given by
\begin{subequations}
\begin{align}
    A^{\mathcal{L}1}_l &= \frac{1}{\sqrt{2\Delta\Sigma}} \Big[i a \sin \theta \,\partial_t - \partial_\theta + i \csc \theta \,\partial_\varphi - \cot \theta \Big] \Big(\zeta \sqrt{\Delta} \Phi^{\mathcal{L}1}_0 \Big), \\
    A^{\mathcal{L}1}_n &= \frac{1}{\sqrt{2\Delta\Sigma}} \Big[i a \sin \theta \,\partial_t + \partial_\theta + i \csc \theta \,\partial_\varphi + \cot \theta \Big] \Big(\zeta \sqrt{\Delta} \Phi^{\mathcal{L}1}_2\Big), \\
    A^{\mathcal{L}1}_m &= \frac{-1}{\Delta\sqrt{2\Sigma}} \Big[(a^2+r^2)\,\partial_t - \Delta \partial_r + a \,\partial_\varphi \Big] \Big(\zeta \sqrt{\Delta} \Phi^{\mathcal{L}1}_0 \Big), \\
    A^{\mathcal{L}1}_\mb &= \frac{1}{\Delta\sqrt{2\Sigma}} \Big[(a^2+r^2)\,\partial_t + \Delta \partial_r + a \,\partial_\varphi\Big] \Big(\zeta \sqrt{\Delta} \Phi^{\mathcal{L}1}_2\Big),
\end{align}
\end{subequations}
and
\begin{subequations}
\begin{align}
    A^{\mathcal{L}2}_l &= \frac{i a \cos \theta}{\sqrt{2\Delta \Sigma}} \Big[i a \sin \theta\,\partial_t - \partial_\theta + i \csc \theta \,\partial_\varphi - \csc \theta  \sec \theta \Big] \Big(\zeta \sqrt{\Delta} \Phi^{\mathcal{L}1}_0 \Big), \\
    A^{\mathcal{L}2}_n &=  \frac{i a \cos \theta}{\sqrt{2\Delta \Sigma}} \Big[-i a \sin \theta\,\partial_t - \partial_\theta - i \csc \theta \,\partial_\varphi - \csc \theta  \sec \theta \Big] \Big(\zeta \sqrt{\Delta} \Phi^{\mathcal{L}1}_2 \Big), \\
    A^{\mathcal{L}2}_m &= \frac{1}{\Delta\sqrt{2\Sigma}} \Big[r (a^2+r^2)\,\partial_t - r\Delta \partial_r + a r \,\partial_\varphi +\Delta \Big] \Big(\zeta \sqrt{\Delta} \Phi^{\mathcal{L}1}_0 \Big), \\
    A^{\mathcal{L}2}_\mb &= \frac{1}{\Delta\sqrt{2\Sigma}} \Big[r (a^2+r^2)\,\partial_t + r\Delta \partial_r + a r \,\partial_\varphi -\Delta \Big] \Big(\zeta \sqrt{\Delta} \Phi^{\mathcal{L}1}_2 \Big).
\end{align}
\end{subequations}
Note the explicitly separated form for the vector potentials. Despite the fact that the vector wave equation does not appear to be
directly separable in Kerr spacetime, its solutions can be written as a separated
operator (up to overall factors of $\sqrt{\Sigma}$ and $\zeta$) acting on a pair of scalar functions that
satisfy separable equations. This separable form for the vector potential is in contrast to the radiation gauge vector potentials which involve a non-separated operator acting on a scalar.

\subsection{Mode decomposed Teukolsky equation}

When working with Carter's tetrad, the Teukolsky equation separates using the ansatz\footnote{The factor of $\Delta^{s/2}$ is included to make ${}_s \phi_{\ell \emm \omega}(r)$ consistent with Teukolsky's function, it is not required to obtain a separable equation.}
\begin{subequations}
\begin{align}
    \phi_0 &=  \frac{\sqrt{\Delta}}{\zeta}\int_{-\infty}^\infty \sum_{\ell=1}^\infty \sum_{\emm=-\ell}^\ell {}_1 \phi_{\ell \emm \omega}(r) {}_1 S_{\ell \emm}(\theta, \varphi; a \omega) e^{-i \omega t} d\omega, \\
    \phi_2 &= \frac{1}{\zeta\sqrt{\Delta}} \int_{-\infty}^\infty \sum_{\ell=1}^\infty \sum_{\emm=-\ell}^\ell {}_{-1} \phi_{\ell \emm \omega}(r) {}_{-1} S_{\ell \emm}(\theta, \varphi; a \omega) e^{-i \omega t} d\omega,
\end{align}
\end{subequations}
with the functions ${}_s \phi_{\ell \emm \omega}(r)$ and ${}_s S_{\ell \emm}(\theta, \varphi; a \omega)$ satisfying the Teukolsky radial and spin-weighted spheroidal harmonic equations,
\begin{subequations}
\begin{align}
  \label{eq:SWSH}
  \bigg[\dfrac{d}{d\chi}\bigg((1-\chi^2)\dfrac{d}{d\chi}\bigg) + a^2 \omega^2 \chi^2 -\frac{(\emm+s \chi)^2}{1-\chi^2} - 2 a s \omega \chi +s + A\bigg]{}_s S_{\ell \emm} &= 0, \\
\label{eq:TeukolskyR}
  \bigg[\Delta^{-s} \dfrac{d}{dr}\bigg(\Delta^{s+1}\dfrac{d}{dr}\bigg) + \frac{K^2 - 2 i s (r-M)K}{\Delta}+ 4 i s \omega r - {}_s \lambda_{\ell \emm} \bigg]{}_s \phi_{\ell \emm \omega} &= {}_s J_{\ell \emm \omega},
\end{align}
\end{subequations}
where $\chi \equiv \cos \theta$, $A \equiv {}_s \lambda_{\ell \emm}+2 a \emm \omega -a^2 \omega^2$ and $K\equiv(r^2+a^2)\omega-a \emm$, and where the
eigenvalue ${}_s \lambda_{\ell \emm}$ depends on the value of $a \omega$. The sources for the radial Teukolsky equation are defined by
\begin{subequations}
\begin{align}
  \mathcal{S}_0 J &= -\frac{\sqrt{\Delta}}{2\Sigma\zeta}\int_{-\infty}^\infty \sum_{\ell=1}^\infty \sum_{\emm=-\ell}^\ell {}_1 J_{\ell \emm \omega}(r) {}_1 S_{\ell \emm}(\theta, \varphi; a \omega) e^{-i \omega t} d\omega, \\
  \mathcal{S}_2 J &= -\frac{1}{2\Sigma\zeta\sqrt{\Delta}} \int_{-\infty}^\infty \sum_{\ell=1}^\infty \sum_{\emm=-\ell}^\ell {}_{-1} J_{\ell \emm \omega}(r) {}_{-1} S_{\ell \emm}(\theta, \varphi; a \omega) e^{-i \omega t} d\omega.
\end{align}
\end{subequations}
Finally, when acting on a single mode of the mode-decomposed Maxwell scalars the symmetry operators yield
\begin{alignat}{8}
  \mathscr{S} \phi_0 &= - \frac12 {}_{|1|} \lambda_{\ell \emm} \phi_0,& \quad
  \mathscr{R} \phi_0 &= - \frac12 {}_{|1|} \lambda_{\ell \emm} \phi_0 + \zeta \bar{\zeta} \mathcal{S}_0 J, &\quad
  \mathscr{S}' \phi_2 &= - \frac12 {}_{|-1|} \lambda_{\ell \emm} \phi_2,& \quad
  \mathscr{R}' \phi_2 &= - \frac12 {}_{|1|} \lambda_{\ell \emm} \phi_2 + \zeta \bar{\zeta} \mathcal{S}_2 J,
\end{alignat}
where ${}_{|s|} \lambda_{\ell \emm \omega} \equiv {}_{s} \lambda_{\ell \emm \omega} + |s| + s$ is independent of the sign of $s$.\footnote{This is distinct from Chandrasekhar's eigenvalue, ${}_{s} \lambda^{\rm Ch}_{\ell \emm \omega} = {}_{s} \lambda_{\ell \emm \omega} + s^2 + s$.}

\end{widetext}

\subsection{Mode decomposed inversion relations for Lorenz-gauge Hertz potentials}

In the homogeneous case, if we decompose the Lorenz-gauge Hertz potentials with the same mode ansatz as for the Maxwell scalars, the inversion relation between modes of the Lorenz-gauge Hertz potentials and the modes of the Maxwell scalars is given by
\begin{alignat}{4}
  {}_{1} \Phi^{\mathcal{L}1}_{\ell\emm\omega} &= - \frac{{}_{1} \phi_{\ell\emm\omega}}{i \omega}, &\quad
  {}_{-1} \Phi^{\mathcal{L}1}_{\ell\emm\omega} &= \frac{{}_{-1} \phi_{\ell\emm\omega}}{i \omega}, \nonumber \\
  {}_{1} \Phi^{\mathcal{L}2}_{\ell\emm\omega} &= -\frac{2 {}_{1} \phi_{\ell\emm\omega}}{{}_{|1|} \lambda_{\ell \emm \omega}}, &\quad
  {}_{-1} \Phi^{\mathcal{L}2}_{\ell\emm\omega} &= \frac{2 {}_{-1} \phi_{\ell\emm\omega}}{{}_{|-1|} \lambda_{\ell \emm \omega}}, \nonumber \\
  {}_{1} \Phi^{\mathcal{L}3}_{\ell\emm\omega} &= \frac{{}_{1} \phi_{\ell\emm\omega}}{(\mathcal{C}_{\ell \emm \omega})^2}, &\quad
  {}_{-1} \Phi^{\mathcal{L}3}_{\ell\emm\omega} &= -\frac{{}_{-1} \phi_{\ell\emm\omega}}{(\mathcal{C}_{\ell \emm \omega})^2}.
\end{alignat}
where $(\mathcal{C}_{\ell \emm \omega})^2 = {}_{|s|} \lambda_{\ell \emm \omega}^2 + 4 \omega a(\emm - a\omega) $ is the (squared) Teukolsky-Starobinsky constant.

\section{Conclusions}
\label{sec:Conclusions}

In this paper we have reviewed and extended recent results for a Hertz potential for the Lorenz gauge Maxwell (vector wave) equation. Dolan \cite{Dolan:2019hcw} previously found a mode version of the potential $\mathcal{L}1$, but our derivation is the first time it has been given without relying on a mode decomposition. The other two potentials, $\mathcal{L}2$ and $\mathcal{L}3$ appear to be new, and have not previously been given in the literature.

There are several important directions for future study. Our work so far has been restricted to the homogenous case
(as was all previous work on Lorenz gauge Hertz potentials). It would be desirable to extend this to allow for sourced
perturbations. It is quite likely that this would be possible with our method by retaining terms
involving the Teukolsky equation that we have eliminated in the final steps of our derivation. It may
also be necessary to apply the ``corrector tensor'' method of Green, Hollands and Zimmerman
\cite{Green:2019nam,Hollands:2020vjg} in order to obtain the most general sourced perturbations. We leave this for future
work.

With an extension of this work to sourced perturbations, this would open the approach up to a number of applications. For example, it would enable the calculation of the Lorenz gauge vector potential sourced by a point electric charge (i.e. the electromagnetic self-force problem) without having to be concerned with the string-like singularities that appear in the radiation gauge. Additionally, it may provide a path to solving the gravitational Lorenz gauge equations by solving for the gauge vector (which satisfies the vector wave equation) required to transform from, e.g., radiation gauge to Lorenz gauge.

A more challenging goal is to extend our work directly to the gravitational case, where we are interested in solving the Lorenz
gauge equation for metric perturbations. Many of the identities (including a Teukolsky equation and
Teukolsky-Starobinsky identities) we have used in the electromagnetic case have analogues in the gravitational case.
One possible complication is that the Teukolsky-Starobinsky identities for the perturbed Weyl scalars mix not only the
scalars $\psi_0$ and $\psi_4$, but also their complex conjugates. We would therefore not expect the simplification we
found using an anti-self-dual bivector to apply in the gravitational case. Nonetheless, there is reason for optimism
that the gravitational case is solvable given the otherwise strong similarities to the electromagnetic case.

\acknowledgements
We thank Adrian Ottewill, Marc Casals, Saul Teukolsky, Amos Ori, Sam Dolan, Leanne Durkan, Stephen Green and Peter Zimmerman
for helpful discussions.
The derivations in this work made extensive use of the \textsc{xAct} \cite{xTensor,xTensorOnline} tensor algebra package for \textsc{Mathematica}.
This work was supported by Enterprise Ireland H2020 Proposal Preparation Support Project CS20182136.

\appendix

\section{Lorenz gauge field equations with Lorenz gauge Hertz potential}
\label{sec:LorenzHertzField}

Here we give an explicit derivation showing that our first potential, $A^{\mathcal{L}1}_\alpha$, satisfies the homogeneous Lorenz gauge field equations (for brevity, we omit similar derivation for the other potentials). Substituting the expression for our potential in terms of $\mathcal{H}_{\alpha\beta}$, Eq.~\eqref{eq:AL1}, into the vacuum vector wave equation, Eq.~\eqref{eq:VectorWave} with $R_{\alpha\beta}=0$, and commuting the wave operator with the covariant derivative, we find that $\mathcal{H}_{ab}$ must satisfy
\begin{equation}
  \label{eq:BoxH-EM}
  \nabla^\nu\big[\Box \zeta \mathcal{H}_{\mu\nu} + 2 R_\mu{}^\alpha{}_\nu{}^\beta \zeta \mathcal{H}_{\alpha \beta}\big] + \nabla^\gamma R_{\mu \alpha \beta \gamma} \zeta \mathcal{H}^{\alpha \beta}=0.
\end{equation}
In type D spacetimes the final term vanishes and the condition reduces to the requirement that $H_{\mu\nu}$ satisfies the divergence of a tensor wave equation. The tetrad components of $A_{\mu\nu} \equiv \Box \zeta \mathcal{H}_{\mu\nu} + 2 R_\mu{}^\alpha{}_\nu{}^\beta \zeta \mathcal{H}_{\alpha \beta}$ are given in GHP form by
\begin{subequations}
\begin{equation}
  A_{l n} - A_{m \bar{m}} = 4 (\rho' \edth'-\tau' \th')\zeta \Phi_0^{\mathcal{L}1} + 4 (\rho \edth - \tau \th) \zeta \Phi_2^{\mathcal{L}1},
\end{equation}
\begin{align}
  A_{l m} &=
    2 \Big[ (\edth - \bar{\tau}') (\edth' - \tau') - (\th - \bar{\rho}) (\th' - \rho') \Big] \zeta \Phi_0^{\mathcal{L}1} \nonumber \\
    &= 2 [\mathcal{\pounds}_\xi - \zeta \mathcal{O}] \Phi_0^{\mathcal{L}1},
\end{align}
\end{subequations}
with all other components given by symmetries, $A_{l n} + A_{m \bar{m}} = \overline{A_{l n} - A_{m \bar{m}}}$, $A_{l \bar{m}} = \overline{A_{l m}}$, $A_{nm} = \overline{A_{lm}'}$ and $A_{n\bar{m}} = A_{lm}'$ along with the identification $\Phi_0^{\mathcal{L}1} = -(\Phi^{\mathcal{L}1}_2)'$.
We now wish to evaluate the Lorenz gauge equation by computing $\nabla^\mu A_{\mu \nu}$. Doing so, we recover Eq.~\eqref{eq:LorenzL1}. Note that this does not rely on any cancellations between $\Phi_0^{\mathcal{L}1}$ and $\Phi_2^{\mathcal{L}1}$ and their complex conjugates; the field equations are satisfied by $A^{\mathcal{L}1}_{a}$ and its complex conjugate independently.

The above conclusion for the Lorenz gauge field equation can also be obtained by an alternative approach. Requiring that Eq.~\eqref{eq:BoxH-EM} is satisfied means that, for a suitable choice of ``gauge vector of third kind", $X^\alpha$, we should be able to solve
\begin{equation}
 \Box \zeta \mathcal{H}_{\mu\nu} + 2 R_\mu{}^\alpha{}_\nu{}^\beta \zeta \mathcal{H}_{\alpha \beta} + \epsilon_{\mu\nu}{}^{\alpha \beta} X_{\alpha;\beta} = 0.
\end{equation}
To solve for $X_\mu$ for our given $\mathcal{H}_{\mu \nu}$ we need only consider the two components
\begin{subequations}
\begin{align}
  \epsilon_{lm}{}^{\gamma \delta} X_{\gamma;\delta} &= i \big[(\th-\bar{\rho}) X_m - (\edth-\bar{\tau}') X_l \big], \\
  \epsilon_{n\bar{m}}{}^{\gamma \delta} X_{\gamma;\delta} &= i \big[(\th'-\bar{\rho}') X_{\bar{m}} - (\edth'-\bar{\tau}) X_n \big],
\end{align}
\end{subequations}
the second of which is the prime of the first. It is clear by inspection that these would cancel $A_{lm}$ and $A_{n\bar{m}}$ by virtue of the Teukolsky equation if
\begin{subequations}
\begin{align}
  X_l &= 2 i \zeta^{-1} \edth' \zeta^2 \Phi_0^{\mathcal{L}1}\\
  X_n &= -2 i \zeta^{-1} \edth \zeta^2 \Phi_2^{\mathcal{L}1}\\
  X_m &= 2 i \zeta^{-1} \th' \zeta^2 \Phi_0^{\mathcal{L}1}\\
  X_\mb &= -2 i \zeta^{-1} \th \zeta^2 \Phi_2^{\mathcal{L}1},
\end{align}
\end{subequations}
where the second and fourth lines are the prime of the first and third lines, respectively.
It is straightforward to check that the remaining components of $A_{\alpha \beta}$ are also reproduced with this $X^\alpha$.

\bibliography{SeparableKerrEM}

\begin{thebibliography}{29}%
\makeatletter
\providecommand \@ifxundefined [1]{%
 \@ifx{#1\undefined}
}%
\providecommand \@ifnum [1]{%
 \ifnum #1\expandafter \@firstoftwo
 \else \expandafter \@secondoftwo
 \fi
}%
\providecommand \@ifx [1]{%
 \ifx #1\expandafter \@firstoftwo
 \else \expandafter \@secondoftwo
 \fi
}%
\providecommand \natexlab [1]{#1}%
\providecommand \enquote  [1]{``#1''}%
\providecommand \bibnamefont  [1]{#1}%
\providecommand \bibfnamefont [1]{#1}%
\providecommand \citenamefont [1]{#1}%
\providecommand \href@noop [0]{\@secondoftwo}%
\providecommand \href [0]{\begingroup \@sanitize@url \@href}%
\providecommand \@href[1]{\@@startlink{#1}\@@href}%
\providecommand \@@href[1]{\endgroup#1\@@endlink}%
\providecommand \@sanitize@url [0]{\catcode `\\12\catcode `\$12\catcode
  `\&12\catcode `\#12\catcode `\^12\catcode `\_12\catcode `\%12\relax}%
\providecommand \@@startlink[1]{}%
\providecommand \@@endlink[0]{}%
\providecommand \url  [0]{\begingroup\@sanitize@url \@url }%
\providecommand \@url [1]{\endgroup\@href {#1}{\urlprefix }}%
\providecommand \urlprefix  [0]{URL }%
\providecommand \Eprint [0]{\href }%
\providecommand \doibase [0]{http://dx.doi.org/}%
\providecommand \selectlanguage [0]{\@gobble}%
\providecommand \bibinfo  [0]{\@secondoftwo}%
\providecommand \bibfield  [0]{\@secondoftwo}%
\providecommand \translation [1]{[#1]}%
\providecommand \BibitemOpen [0]{}%
\providecommand \bibitemStop [0]{}%
\providecommand \bibitemNoStop [0]{.\EOS\space}%
\providecommand \EOS [0]{\spacefactor3000\relax}%
\providecommand \BibitemShut  [1]{\csname bibitem#1\endcsname}%
\let\auto@bib@innerbib\@empty
%</preamble>
\bibitem [{\citenamefont {Teukolsky}(1972)}]{Teukolsky:1972my}%
  \BibitemOpen
  \bibfield  {author} {\bibinfo {author} {\bibfnamefont {Saul~A.}\ \bibnamefont
  {Teukolsky}},\ }\bibfield  {title} {\enquote {\bibinfo {title} {{Rotating
  black holes - separable wave equations for gravitational and electromagnetic
  perturbations}},}\ }\href {\doibase 10.1103/PhysRevLett.29.1114} {\bibfield
  {journal} {\bibinfo  {journal} {Phys. Rev. Lett.}\ }\textbf {\bibinfo
  {volume} {29}},\ \bibinfo {pages} {1114--1118} (\bibinfo {year}
  {1972})}\BibitemShut {NoStop}%
%%CITATION = PRLTA,29,1114;%%
\bibitem [{\citenamefont {Chrzanowski}(1975)}]{Chrzanowski:1975wv}%
  \BibitemOpen
  \bibfield  {author} {\bibinfo {author} {\bibfnamefont {P.~L.}\ \bibnamefont
  {Chrzanowski}},\ }\bibfield  {title} {\enquote {\bibinfo {title} {{Vector
  Potential and Metric Perturbations of a Rotating Black Hole}},}\ }\href
  {\doibase 10.1103/PhysRevD.11.2042} {\bibfield  {journal} {\bibinfo
  {journal} {Phys. Rev.}\ }\textbf {\bibinfo {volume} {D11}},\ \bibinfo {pages}
  {2042--2062} (\bibinfo {year} {1975})}\BibitemShut {NoStop}%
%%CITATION = PHRVA,D11,2042;%%
\bibitem [{\citenamefont {Kegeles}\ and\ \citenamefont
  {Cohen}(1979)}]{Kegeles:1979an}%
  \BibitemOpen
  \bibfield  {author} {\bibinfo {author} {\bibfnamefont {L.~S.}\ \bibnamefont
  {Kegeles}}\ and\ \bibinfo {author} {\bibfnamefont {J.~M.}\ \bibnamefont
  {Cohen}},\ }\bibfield  {title} {\enquote {\bibinfo {title} {{Constructive
  procedure for perturbations of space-times}},}\ }\href {\doibase
  10.1103/PhysRevD.19.1641} {\bibfield  {journal} {\bibinfo  {journal} {Phys.
  Rev.}\ }\textbf {\bibinfo {volume} {D19}},\ \bibinfo {pages} {1641--1664}
  (\bibinfo {year} {1979})}\BibitemShut {NoStop}%
%%CITATION = PHRVA,D19,1641;%%
\bibitem [{\citenamefont {Wald}(1978)}]{Wald:1978vm}%
  \BibitemOpen
  \bibfield  {author} {\bibinfo {author} {\bibfnamefont {Robert~M.}\
  \bibnamefont {Wald}},\ }\bibfield  {title} {\enquote {\bibinfo {title}
  {{Construction of Solutions of Gravitational, Electromagnetic, Or Other
  Perturbation Equations from Solutions of Decoupled Equations}},}\ }\href
  {\doibase 10.1103/PhysRevLett.41.203} {\bibfield  {journal} {\bibinfo
  {journal} {Phys. Rev. Lett.}\ }\textbf {\bibinfo {volume} {41}},\ \bibinfo
  {pages} {203--206} (\bibinfo {year} {1978})}\BibitemShut {NoStop}%
%%CITATION = PRLTA,41,203;%%
\bibitem [{\citenamefont {Whiting}\ and\ \citenamefont
  {Price}(2005)}]{Whiting:2005hr}%
  \BibitemOpen
  \bibfield  {author} {\bibinfo {author} {\bibfnamefont {B.F.}\ \bibnamefont
  {Whiting}}\ and\ \bibinfo {author} {\bibfnamefont {L.R.}\ \bibnamefont
  {Price}},\ }\bibfield  {title} {\enquote {\bibinfo {title} {{Metric
  reconstruction from Weyl scalars}},}\ }\href {\doibase
  10.1088/0264-9381/22/15/003} {\bibfield  {journal} {\bibinfo  {journal}
  {Class. Quant. Grav.}\ }\textbf {\bibinfo {volume} {22}},\ \bibinfo {pages}
  {S589--S604} (\bibinfo {year} {2005})}\BibitemShut {NoStop}%
\bibitem [{\citenamefont {Pound}\ \emph {et~al.}(2014)\citenamefont {Pound},
  \citenamefont {Merlin},\ and\ \citenamefont {Barack}}]{Pound:2013faa}%
  \BibitemOpen
  \bibfield  {author} {\bibinfo {author} {\bibfnamefont {Adam}\ \bibnamefont
  {Pound}}, \bibinfo {author} {\bibfnamefont {Cesar}\ \bibnamefont {Merlin}}, \
  and\ \bibinfo {author} {\bibfnamefont {Leor}\ \bibnamefont {Barack}},\
  }\bibfield  {title} {\enquote {\bibinfo {title} {{Gravitational self-force
  from radiation-gauge metric perturbations}},}\ }\href {\doibase
  10.1103/PhysRevD.89.024009} {\bibfield  {journal} {\bibinfo  {journal} {Phys.
  Rev.}\ }\textbf {\bibinfo {volume} {D89}},\ \bibinfo {pages} {024009}
  (\bibinfo {year} {2014})},\ \Eprint {http://arxiv.org/abs/1310.1513}
  {arXiv:1310.1513 [gr-qc]} \BibitemShut {NoStop}%
%%CITATION = ARXIV:1310.1513;%%
\bibitem [{\citenamefont {Stewart}(1979)}]{Stewart:1978tm}%
  \BibitemOpen
  \bibfield  {author} {\bibinfo {author} {\bibfnamefont {John~M.}\ \bibnamefont
  {Stewart}},\ }\bibfield  {title} {\enquote {\bibinfo {title}
  {{{Hertz-Bromwich-Debye-Whittaker-Penrose} Potentials in General
  Relativity}},}\ }\href {\doibase 10.1098/rspa.1979.0101} {\bibfield
  {journal} {\bibinfo  {journal} {Proc. Roy. Soc. Lond.}\ }\textbf {\bibinfo
  {volume} {A367}},\ \bibinfo {pages} {527--538} (\bibinfo {year}
  {1979})}\BibitemShut {NoStop}%
%%CITATION = PRSLA,A367,527;%%
\bibitem [{\citenamefont {Sago}\ \emph {et~al.}(2003)\citenamefont {Sago},
  \citenamefont {Nakano},\ and\ \citenamefont {Sasaki}}]{Sago:2002fe}%
  \BibitemOpen
  \bibfield  {author} {\bibinfo {author} {\bibfnamefont {Norichika}\
  \bibnamefont {Sago}}, \bibinfo {author} {\bibfnamefont {Hiroyuki}\
  \bibnamefont {Nakano}}, \ and\ \bibinfo {author} {\bibfnamefont {Misao}\
  \bibnamefont {Sasaki}},\ }\bibfield  {title} {\enquote {\bibinfo {title}
  {{Gauge problem in the gravitational selfforce. 1. Harmonic gauge approach in
  the Schwarzschild background}},}\ }\href {\doibase
  10.1103/PhysRevD.67.104017} {\bibfield  {journal} {\bibinfo  {journal} {Phys.
  Rev. D}\ }\textbf {\bibinfo {volume} {67}},\ \bibinfo {pages} {104017}
  (\bibinfo {year} {2003})},\ \Eprint {http://arxiv.org/abs/gr-qc/0208060}
  {arXiv:gr-qc/0208060} \BibitemShut {NoStop}%
\bibitem [{\citenamefont {Lunin}(2017)}]{Lunin:2017drx}%
  \BibitemOpen
  \bibfield  {author} {\bibinfo {author} {\bibfnamefont {Oleg}\ \bibnamefont
  {Lunin}},\ }\bibfield  {title} {\enquote {\bibinfo {title}
  {{Maxwell\textquoteright{}s equations in the Myers-Perry geometry}},}\ }\href
  {\doibase 10.1007/JHEP12(2017)138} {\bibfield  {journal} {\bibinfo  {journal}
  {JHEP}\ }\textbf {\bibinfo {volume} {12}},\ \bibinfo {pages} {138} (\bibinfo
  {year} {2017})},\ \Eprint {http://arxiv.org/abs/1708.06766} {arXiv:1708.06766
  [hep-th]} \BibitemShut {NoStop}%
\bibitem [{\citenamefont {Frolov}\ \emph
  {et~al.}(2018{\natexlab{a}})\citenamefont {Frolov}, \citenamefont
  {Krtou\v{s}},\ and\ \citenamefont {Kubiz\v{n}\'ak}}]{Frolov:2018pys}%
  \BibitemOpen
  \bibfield  {author} {\bibinfo {author} {\bibfnamefont {Valeri~P.}\
  \bibnamefont {Frolov}}, \bibinfo {author} {\bibfnamefont {Pavel}\
  \bibnamefont {Krtou\v{s}}}, \ and\ \bibinfo {author} {\bibfnamefont {David}\
  \bibnamefont {Kubiz\v{n}\'ak}},\ }\bibfield  {title} {\enquote {\bibinfo
  {title} {{Separation of variables in Maxwell equations in
  Pleba\'nski-Demia\'nski spacetime}},}\ }\href {\doibase
  10.1103/PhysRevD.97.101701} {\bibfield  {journal} {\bibinfo  {journal} {Phys.
  Rev. D}\ }\textbf {\bibinfo {volume} {97}},\ \bibinfo {pages} {101701}
  (\bibinfo {year} {2018}{\natexlab{a}})},\ \Eprint
  {http://arxiv.org/abs/1802.09491} {arXiv:1802.09491 [hep-th]} \BibitemShut
  {NoStop}%
\bibitem [{\citenamefont {Krtou\v{s}}\ \emph {et~al.}(2018)\citenamefont
  {Krtou\v{s}}, \citenamefont {Frolov},\ and\ \citenamefont
  {Kubiz\v{n}\'ak}}]{Krtous:2018bvk}%
  \BibitemOpen
  \bibfield  {author} {\bibinfo {author} {\bibfnamefont {Pavel}\ \bibnamefont
  {Krtou\v{s}}}, \bibinfo {author} {\bibfnamefont {Valeri~P.}\ \bibnamefont
  {Frolov}}, \ and\ \bibinfo {author} {\bibfnamefont {David}\ \bibnamefont
  {Kubiz\v{n}\'ak}},\ }\bibfield  {title} {\enquote {\bibinfo {title}
  {{Separation of Maxwell equations in Kerr\textendash{}NUT\textendash{}(A)dS
  spacetimes}},}\ }\href {\doibase 10.1016/j.nuclphysb.2018.06.019} {\bibfield
  {journal} {\bibinfo  {journal} {Nucl. Phys. B}\ }\textbf {\bibinfo {volume}
  {934}},\ \bibinfo {pages} {7--38} (\bibinfo {year} {2018})},\ \Eprint
  {http://arxiv.org/abs/1803.02485} {arXiv:1803.02485 [hep-th]} \BibitemShut
  {NoStop}%
\bibitem [{\citenamefont {Frolov}\ \emph
  {et~al.}(2018{\natexlab{b}})\citenamefont {Frolov}, \citenamefont
  {Krtou\v{s}}, \citenamefont {Kubiz\v{n}\'ak},\ and\ \citenamefont
  {Santos}}]{Frolov:2018ezx}%
  \BibitemOpen
  \bibfield  {author} {\bibinfo {author} {\bibfnamefont {Valeri~P.}\
  \bibnamefont {Frolov}}, \bibinfo {author} {\bibfnamefont {Pavel}\
  \bibnamefont {Krtou\v{s}}}, \bibinfo {author} {\bibfnamefont {David}\
  \bibnamefont {Kubiz\v{n}\'ak}}, \ and\ \bibinfo {author} {\bibfnamefont
  {Jorge~E.}\ \bibnamefont {Santos}},\ }\bibfield  {title} {\enquote {\bibinfo
  {title} {{Massive Vector Fields in Rotating Black-Hole Spacetimes:
  Separability and Quasinormal Modes}},}\ }\href {\doibase
  10.1103/PhysRevLett.120.231103} {\bibfield  {journal} {\bibinfo  {journal}
  {Phys. Rev. Lett.}\ }\textbf {\bibinfo {volume} {120}},\ \bibinfo {pages}
  {231103} (\bibinfo {year} {2018}{\natexlab{b}})},\ \Eprint
  {http://arxiv.org/abs/1804.00030} {arXiv:1804.00030 [hep-th]} \BibitemShut
  {NoStop}%
\bibitem [{\citenamefont {Dolan}(2018)}]{Dolan:2018dqv}%
  \BibitemOpen
  \bibfield  {author} {\bibinfo {author} {\bibfnamefont {Sam~R.}\ \bibnamefont
  {Dolan}},\ }\bibfield  {title} {\enquote {\bibinfo {title} {{Instability of
  the Proca field on Kerr spacetime}},}\ }\href {\doibase
  10.1103/PhysRevD.98.104006} {\bibfield  {journal} {\bibinfo  {journal} {Phys.
  Rev. D}\ }\textbf {\bibinfo {volume} {98}},\ \bibinfo {pages} {104006}
  (\bibinfo {year} {2018})},\ \Eprint {http://arxiv.org/abs/1806.01604}
  {arXiv:1806.01604 [gr-qc]} \BibitemShut {NoStop}%
\bibitem [{\citenamefont {Dolan}(2019)}]{Dolan:2019hcw}%
  \BibitemOpen
  \bibfield  {author} {\bibinfo {author} {\bibfnamefont {Sam~R.}\ \bibnamefont
  {Dolan}},\ }\bibfield  {title} {\enquote {\bibinfo {title} {{Electromagnetic
  fields on Kerr spacetime, Hertz potentials and Lorenz gauge}},}\ }\href
  {\doibase 10.1103/PhysRevD.100.044044} {\bibfield  {journal} {\bibinfo
  {journal} {Phys. Rev. D}\ }\textbf {\bibinfo {volume} {100}},\ \bibinfo
  {pages} {044044} (\bibinfo {year} {2019})},\ \Eprint
  {http://arxiv.org/abs/1906.04808} {arXiv:1906.04808 [gr-qc]} \BibitemShut
  {NoStop}%
\bibitem [{\citenamefont {Houri}\ \emph {et~al.}(2020)\citenamefont {Houri},
  \citenamefont {Tanahashi},\ and\ \citenamefont {Yasui}}]{Houri:2019lnu}%
  \BibitemOpen
  \bibfield  {author} {\bibinfo {author} {\bibfnamefont {Tsuyoshi}\
  \bibnamefont {Houri}}, \bibinfo {author} {\bibfnamefont {Norihiro}\
  \bibnamefont {Tanahashi}}, \ and\ \bibinfo {author} {\bibfnamefont
  {Yukinori}\ \bibnamefont {Yasui}},\ }\bibfield  {title} {\enquote {\bibinfo
  {title} {{On symmetry operators for the Maxwell equation on the
  Kerr-NUT-(A)dS spacetime}},}\ }\href {\doibase 10.1088/1361-6382/ab586d}
  {\bibfield  {journal} {\bibinfo  {journal} {Class. Quant. Grav.}\ }\textbf
  {\bibinfo {volume} {37}},\ \bibinfo {pages} {015011} (\bibinfo {year}
  {2020})},\ \Eprint {http://arxiv.org/abs/1908.10250} {arXiv:1908.10250
  [gr-qc]} \BibitemShut {NoStop}%
\bibitem [{\citenamefont {Misner}\ \emph {et~al.}(1973)\citenamefont {Misner},
  \citenamefont {Thorne},\ and\ \citenamefont {Wheeler}}]{Misner:1974qy}%
  \BibitemOpen
  \bibfield  {author} {\bibinfo {author} {\bibfnamefont {Charles~W.}\
  \bibnamefont {Misner}}, \bibinfo {author} {\bibfnamefont {K.S.}\ \bibnamefont
  {Thorne}}, \ and\ \bibinfo {author} {\bibfnamefont {J.A.}\ \bibnamefont
  {Wheeler}},\ }\href@noop {} {\emph {\bibinfo {title} {{Gravitation}}}}\
  (\bibinfo  {publisher} {{Freeman}},\ \bibinfo {address} {San Francisco},\
  \bibinfo {year} {1973})\BibitemShut {NoStop}%
%%CITATION = ISBN-9780716703440 ETC.;%%
\bibitem [{\citenamefont {Geroch}\ \emph {et~al.}(1973)\citenamefont {Geroch},
  \citenamefont {Held},\ and\ \citenamefont {Penrose}}]{Geroch:1973am}%
  \BibitemOpen
  \bibfield  {author} {\bibinfo {author} {\bibfnamefont {Robert~P.}\
  \bibnamefont {Geroch}}, \bibinfo {author} {\bibfnamefont {A.}~\bibnamefont
  {Held}}, \ and\ \bibinfo {author} {\bibfnamefont {R.}~\bibnamefont
  {Penrose}},\ }\bibfield  {title} {\enquote {\bibinfo {title} {{A space-time
  calculus based on pairs of null directions}},}\ }\href {\doibase
  10.1063/1.1666410} {\bibfield  {journal} {\bibinfo  {journal} {J. Math.
  Phys.}\ }\textbf {\bibinfo {volume} {14}},\ \bibinfo {pages} {874--881}
  (\bibinfo {year} {1973})}\BibitemShut {NoStop}%
%%CITATION = JMAPA,14,874;%%
\bibitem [{\citenamefont {Price}(2007)}]{Price:Thesis}%
  \BibitemOpen
  \bibfield  {author} {\bibinfo {author} {\bibfnamefont {Larry}\ \bibnamefont
  {Price}},\ }\href {http://etd.fcla.edu/UF/UFE0021314/price_l.pdf} {Ph.D.
  thesis},\ \bibinfo  {school} {University of Florida} (\bibinfo {year}
  {2007})\BibitemShut {NoStop}%
\bibitem [{\citenamefont {Aksteiner}(2014)}]{Aksteiner:2014zyp}%
  \BibitemOpen
  \bibfield  {author} {\bibinfo {author} {\bibfnamefont {Steffen}\ \bibnamefont
  {Aksteiner}},\ }\href {\doibase 10.15488/8214} {Ph.D. thesis},\ \bibinfo
  {school} {Leibniz U., Hannover} (\bibinfo {year} {2014})\BibitemShut
  {NoStop}%
\bibitem [{\citenamefont {Penrose}\ and\ \citenamefont
  {Rindler}(2011)}]{Penrose:1987uia}%
  \BibitemOpen
  \bibfield  {author} {\bibinfo {author} {\bibfnamefont {Roger}\ \bibnamefont
  {Penrose}}\ and\ \bibinfo {author} {\bibfnamefont {Wolfgang}\ \bibnamefont
  {Rindler}},\ }\href {\doibase 10.1017/CBO9780511564048} {\emph {\bibinfo
  {title} {{Spinors and Space-Time}}}},\ Cambridge Monographs on Mathematical
  Physics\ (\bibinfo  {publisher} {Cambridge Univ. Press},\ \bibinfo {address}
  {Cambridge, UK},\ \bibinfo {year} {2011})\BibitemShut {NoStop}%
\bibitem [{\citenamefont {Pound}\ and\ \citenamefont
  {Wardell}(2020)}]{Wardell:Pound}%
  \BibitemOpen
  \bibfield  {author} {\bibinfo {author} {\bibfnamefont {Adam}\ \bibnamefont
  {Pound}}\ and\ \bibinfo {author} {\bibfnamefont {Barry}\ \bibnamefont
  {Wardell}},\ }\bibfield  {title} {\enquote {\bibinfo {title} {Black hole
  perturbation theory and gravitational self-force},}\ }in\ \href@noop {}
  {\emph {\bibinfo {booktitle} {Handbook of Gravitational Wave Astronomy}}},\
  \bibinfo {series and number} {Springer Reference},\ \bibinfo {editor} {edited
  by\ \bibinfo {editor} {\bibfnamefont {Cosimo}\ \bibnamefont {Bambi}},
  \bibinfo {editor} {\bibfnamefont {Stavros}\ \bibnamefont {Katsanevas}}, \
  and\ \bibinfo {editor} {\bibfnamefont {Kostantinos~D.}\ \bibnamefont
  {Kokkotas}}}\ (\bibinfo  {publisher} {Springer International Publishing},\
  \bibinfo {year} {2020})\BibitemShut {NoStop}%
\bibitem [{\citenamefont {Fackerell}\ and\ \citenamefont
  {Ipser}(1972)}]{Fackerell:1972hg}%
  \BibitemOpen
  \bibfield  {author} {\bibinfo {author} {\bibfnamefont {E.D.}\ \bibnamefont
  {Fackerell}}\ and\ \bibinfo {author} {\bibfnamefont {J.R.}\ \bibnamefont
  {Ipser}},\ }\bibfield  {title} {\enquote {\bibinfo {title} {{Weak
  electromagnetic fields around a rotating black hole}},}\ }\href {\doibase
  10.1103/PhysRevD.5.2455} {\bibfield  {journal} {\bibinfo  {journal} {Phys.
  Rev. D}\ }\textbf {\bibinfo {volume} {5}},\ \bibinfo {pages} {2455--2458}
  (\bibinfo {year} {1972})}\BibitemShut {NoStop}%
\bibitem [{\citenamefont {Newman}\ \emph {et~al.}(1963)\citenamefont {Newman},
  \citenamefont {Tamburino},\ and\ \citenamefont {Unti}}]{Newman:1963yy}%
  \BibitemOpen
  \bibfield  {author} {\bibinfo {author} {\bibfnamefont {E.}~\bibnamefont
  {Newman}}, \bibinfo {author} {\bibfnamefont {L.}~\bibnamefont {Tamburino}}, \
  and\ \bibinfo {author} {\bibfnamefont {T.}~\bibnamefont {Unti}},\ }\bibfield
  {title} {\enquote {\bibinfo {title} {{Empty space generalization of the
  Schwarzschild metric}},}\ }\href {\doibase 10.1063/1.1704018} {\bibfield
  {journal} {\bibinfo  {journal} {J. Math. Phys.}\ }\textbf {\bibinfo {volume}
  {4}},\ \bibinfo {pages} {915} (\bibinfo {year} {1963})}\BibitemShut {NoStop}%
\bibitem [{\citenamefont {Kinnersley}(1969)}]{Kinnersley:1969zza}%
  \BibitemOpen
  \bibfield  {author} {\bibinfo {author} {\bibfnamefont {William}\ \bibnamefont
  {Kinnersley}},\ }\bibfield  {title} {\enquote {\bibinfo {title} {{Type D
  Vacuum Metrics}},}\ }\href {\doibase 10.1063/1.1664958} {\bibfield  {journal}
  {\bibinfo  {journal} {J. Math. Phys.}\ }\textbf {\bibinfo {volume} {10}},\
  \bibinfo {pages} {1195--1203} (\bibinfo {year} {1969})}\BibitemShut {NoStop}%
%%CITATION = JMAPA,10,1195;%%
\bibitem [{\citenamefont {Carter}\ and\ \citenamefont
  {Hartle}(1987)}]{Carter:1987hk}%
  \BibitemOpen
  \bibfield  {author} {\bibinfo {author} {\bibfnamefont {B.}~\bibnamefont
  {Carter}}\ and\ \bibinfo {author} {\bibfnamefont {J.~B.}\ \bibnamefont
  {Hartle}},\ }\bibfield  {title} {\enquote {\bibinfo {title} {{Gravitation in
  Astrophysics}},}\ }\href {\doibase 10.1007/978-1-4613-1897-2} {\bibfield
  {journal} {\bibinfo  {journal} {NATO Sci. Ser. B}\ }\textbf {\bibinfo
  {volume} {156}},\ \bibinfo {pages} {pp.1--399} (\bibinfo {year}
  {1987})}\BibitemShut {NoStop}%
%%CITATION = INSPIRE-255771;%%
\bibitem [{\citenamefont {Green}\ \emph {et~al.}(2020)\citenamefont {Green},
  \citenamefont {Hollands},\ and\ \citenamefont {Zimmerman}}]{Green:2019nam}%
  \BibitemOpen
  \bibfield  {author} {\bibinfo {author} {\bibfnamefont {Stephen~R.}\
  \bibnamefont {Green}}, \bibinfo {author} {\bibfnamefont {Stefan}\
  \bibnamefont {Hollands}}, \ and\ \bibinfo {author} {\bibfnamefont {Peter}\
  \bibnamefont {Zimmerman}},\ }\bibfield  {title} {\enquote {\bibinfo {title}
  {{Teukolsky formalism for nonlinear Kerr perturbations}},}\ }\href {\doibase
  10.1088/1361-6382/ab7075} {\bibfield  {journal} {\bibinfo  {journal} {Class.
  Quant. Grav.}\ }\textbf {\bibinfo {volume} {37}},\ \bibinfo {pages} {075001}
  (\bibinfo {year} {2020})},\ \Eprint {http://arxiv.org/abs/1908.09095}
  {arXiv:1908.09095 [gr-qc]} \BibitemShut {NoStop}%
\bibitem [{\citenamefont {Hollands}\ and\ \citenamefont
  {Toomani}(2020)}]{Hollands:2020vjg}%
  \BibitemOpen
  \bibfield  {author} {\bibinfo {author} {\bibfnamefont {Stefan}\ \bibnamefont
  {Hollands}}\ and\ \bibinfo {author} {\bibfnamefont {Vahid}\ \bibnamefont
  {Toomani}},\ }\bibfield  {title} {\enquote {\bibinfo {title} {{On the
  radiation gauge for spin-1 perturbations in Kerr-Newman spacetime}},}\
  }\href@noop {} {\  (\bibinfo {year} {2020})},\ \Eprint
  {http://arxiv.org/abs/2008.08550} {arXiv:2008.08550 [gr-qc]} \BibitemShut
  {NoStop}%
\bibitem [{\citenamefont {Mart\'in-Garc\'ia}(2008)}]{xTensor}%
  \BibitemOpen
  \bibfield  {author} {\bibinfo {author} {\bibfnamefont {Jose~M.}\ \bibnamefont
  {Mart\'in-Garc\'ia}},\ }\bibfield  {title} {\enquote {\bibinfo {title}
  {xperm: fast index canonicalization for tensor computer algebra},}\
  }\href@noop {} {\bibfield  {journal} {\bibinfo  {journal} {Comp. Phys.
  Commun.}\ }\textbf {\bibinfo {volume} {179}},\ \bibinfo {pages} {597--603}
  (\bibinfo {year} {2008})}\BibitemShut {NoStop}%
\bibitem [{\citenamefont {Martin-Garcia}()}]{xTensorOnline}%
  \BibitemOpen
  \bibfield  {author} {\bibinfo {author} {\bibfnamefont {J.M.}\ \bibnamefont
  {Martin-Garcia}},\ }\href@noop {} {\enquote {\bibinfo {title} {{xAct:
  Efficient Tensor Computer Algebra for Mathematica}},}\ }\bibinfo
  {howpublished} {\url{http://xact.es/}}\BibitemShut {NoStop}%
\end{thebibliography}%

\end{document}